\pdfoutput=1
\documentclass{elsart3}

\usepackage{graphicx}
\usepackage{cite}
\usepackage{array}
\usepackage{upgreek}    
\usepackage{amssymb}    
\usepackage[figuresright]{rotating}

\newcommand{\beq}{\begin{equation}}
\newcommand{\eeq}{\end{equation}}

\newcommand{\eVdist}{\kern-0.06667em}

\newcommand{\gev}{{\,\textrm{Ge}\eVdist\textrm{V}}}
\newcommand{\mum}{\,\upmu\textrm{m}}
\newcommand{\cm}{\,\textrm{cm}}
\newcommand{\mm}{\,\textrm{mm}}

\newcommand{\ns}{\,\textrm{ns}}
\newcommand{\ms}{\,\textrm{ms}}
\newcommand{\pc}{\,\textrm{\%}}
\newcommand{\mhz}{\,\textrm{MHz}}
\newcommand{\hz}{\,\textrm{Hz}}

\chardef\til=126

\newcommand{\AmS}{{\protect\the\textfont2
  A\kern-.1667em\lower.5ex\hbox{M}\kern-.125emS}}

\begin{document}

\begin{frontmatter}

\title{The design and performance of the ZEUS Micro Vertex Detector}

\author{A.~Polini}
\address{\it University and INFN Bologna, Bologna, Italy\thanksref{bologna}}
\author{I.~Brock}, 
\author{S.~Goers},
\author{A.~Kappes},
\author{U.~F.~Katz}, 
\author{E.~Hilger},
\author{J.~Rautenberg}, 
\author{A.~Weber}
\address{\it Physikalisches Institut der Universit\"at Bonn, Bonn,
Germany\thanksref{bonn}}
\author{A.~Mastroberardino},
\author{E.~Tassi}
\address{\it Calabria University, Physics Department and INFN, Cosenza, Italy\thanksref{bologna}}
\author{V.~Adler}, 
\author{L.~A.~T.~Bauerdick}, 
\author{I.~Bloch}, 
\author{T.~Haas\corauthref{haas}}, 
\author{U.~Klein}, 
\author{U.~Koetz},
\author{G.~Kramberger}, 
\author{E.~Lobodzinska}, 
\author{R.~Mankel},
\author{J.~Ng},
\author{D.~ Notz},
\author{M.~C.~Petrucci}, 
\author{B.~Surrow},
\author{G.~Watt}, 
\author{C.~Youngman}, 
\author{W.~Zeuner}
\address{\it Deutsches Elektronen-Synchrotron DESY, Hamburg, Germany}
\author{C.~Coldewey}, 
\author{R.~Heller}
\address{\it Deutsches Elektronen-Synchrotron DESY, Zeuthen, Germany}
\author{E.~Gallo}
\address{\it University and INFN, Florence, Italy\thanksref{bologna}}
\author{T.~Carli}, 
\author{V.~Chiochia}, 
\author{D.~Dannheim}, 
\author{E.~Fretwurst},
\author{A.~Garfagnini}, 
\author{R.~Klanner,}
\author{B.~Koppitz}, 
\author{J.~Martens}, 
\author{M.~Milite}
\address{\it Hamburg University, Institute of Exp. Physics, Hamburg,
Germany\thanksref{hamburg}}
\author{Katsuo Tokushuku}
\address{\it Institute of Particle and Nuclear Studies, KEK, Tsukuba,
Japan\thanksref{kek}}
\author{I.~Redondo}
\address{\it Universidad Autonoma de Madrid, Madrid, Spain\thanksref{madrid}}
\author{H.~Boterenbrood},
\author{E.~Koffeman}, 
\author{P.~Kooijman}, 
\author{E.~Maddox}, 
\author{H.~Tiecke}, 
\author{M.~Vazquez},
\author{J.~Velthuis},
\author{L.~Wiggers},
\address{\it NIKHEF and University of Amsterdam, Amsterdam, Netherlands\thanksref{nikhef}}
\author{R.C.E.~Devenish}, 
\author{M.~Dawson}, 
\author{J.~Ferrando}, 
\author{G.~Grzelak},
\author{K.~Korcsak-Gorzo},
\author{T.~Matsushita}, 
\author{K.~Oliver}, 
\author{P.~Shield}, 
\author{R.~Walczak}
\address{\it Department of Physics, University of Oxford, Oxford, United Kingdom\thanksref{oxford}}
\newpage
\author{A.~Bertolin},
\author{E.~Borsato},
\author{R.~Carlin}, 
\author{F.~Dal~Corso}, 
\author{A.~Longhin}, 
\author{M.~Turcato}
\address{\it Dipartimento di Fisica dell' Universit\`a and INFN,
Padova, Italy\thanksref{bologna}}
\author{T.~Fusayasu}, 
\author{R.~Hori},
\author{T.~Kohno}, 
\author{S.~Shimizu}
\address{\it Department of Physics, University of Tokyo, Tokyo, Japan\thanksref{kek}}
\author{H.~E.~Larsen},
\author{R.~Sacchi},
\author{A.~Staiano}
\address{\it Universit\`a di Torino and INFN, Torino, Italy\thanksref{bologna}}
\author{M.~Arneodo}, 
\author{M.~Ruspa}
\address{\it Universit\`a del Piemonte Orientale, Novara, and INFN,
Torino, Italy\thanksref{bologna}}
\author{J.~Butterworth}, 
\author{C.~Gwenlan}, 
\author{J.~Fraser}, 
\author{D.~Hayes}, 
\author{M.~Hayes},
\author{J.~Lane}, 
\author{G.~Nixon}, 
\author{M.~Postranecky}, 
\author{M.~Sutton}, 
\author{M.~Warren}
\address{\it Physics and Astronomy Department, University College
London, London, United Kingdom\thanksref{oxford}}

\corauth[haas]{Corresponding author. Deutsches Elektronensynchrotron, Notkestrasse 85, D-22607 Hamburg, Germany, Tel.: +49-40-89983281; email: \texttt{tobias.haas@desy.de}}
\thanks[bologna]{supported by the Italian National Institute for Nuclear Physics (INFN)}
\thanks[bonn]{supported by the German Federal Ministry for Education and Research (BMBF), under contract number No 05 HZ6PDA}
\thanks[hamburg]{supported by the German Federal Ministry for Education and Research (BMBF), under contract number No 05 HZ4GUA}
\thanks[kek]{supported by the Japanese Ministry of Education, Culture, Sports, Science and Technology}
\thanks[madrid]{supported by the Spanish Ministry of Education and Science through funds provided by CICYT}
\thanks[nikhef]{supported by the Netherlands Foundation for Research on Matter (FOM)}
\thanks[oxford]{supported by the Science and Technology Facilities Council, UK}

\begin{abstract}
In order to extend the tracking acceptance, to improve the primary and
secondary vertex reconstruction and thus enhancing the tagging
capabilities for short lived particles, the ZEUS experiment at the HERA
Collider at DESY installed a silicon strip vertex detector. The
barrel part of the detector is a 63~cm long cylinder with silicon
sensors arranged around an elliptical beampipe. The forward part
consists of four circular shaped disks. In total just over 200k
channels are read out using $2.9~{\rm m^2}$ of silicon. In this report
a detailed overview of the design and construction of the
detector is given and the performance of the completed system is
reviewed.
\end{abstract}
\end{frontmatter}

\section{Introduction}
\label{sec:intro}

In the HERA storage ring $27.5\,$GeV positrons or electrons collide
with $920~{\rm GeV}$ protons. The ZEUS experiment
\cite{zeus:1993:bluebook,pl:b293:465} is one of two large
multi-purpose collider detectors installed at HERA. From 1992 to 2000
HERA delivered ca. $200\,{\rm pb^{-1}}$ of luminosity and the ZEUS
experiment recorded ca. 180 million collision events. Following the
successful completion of that program in 2000 the accelerator and the
experiments underwent an extensive upgrade during 9 months in
2000/2001 with the goal to achieve a five-fold increase
in integrated luminosity by 2006.

As part of the upgrade of the ZEUS experiment, the tracking
capabilities were enhanced by the installation of a Microvertex
Detector (MVD) inside the existing Central Tracking Detector
(CTD). The goal was to provide the experiment with the capability of
tagging heavy quarks by identifying displaced vertices.  In addition
the aim was to improve the efficiency, acceptance and resolution of
the tracking system. During the design phase the following
requirements were specified in order to achieve that goal:
\begin{itemize}
\item
Polar angular coverage between $10^\circ - 170^\circ$;

\item
Three spatial measurements per track, in two projections;

\item
$20\mum$ intrinsic hit position resolution for normal incident tracks;

\item
Impact parameter resolution of about $100\mum$ for tracks with polar angle
of $90^\circ$, increasing gradually to $1\,$mm at $20^\circ$, for
tracks with momenta greater than $2\gev$;

\item
Noise occupancy $<10^{-3}$;

\item
Hit efficiency $>97\%$;

\item
Alignment accuracy better than $20\mum$;

\item
Two-track separation better than $200\mum$.
\end{itemize}
A number of constraints also had to be taken into account. These were
\begin{itemize}
\item
The space available for the Microvertex Detector is limited by the
inner bore of the central tracking detector (diameter $324\mm$);

\item
The longitudinal size of the HERA beam spot extends over about $40\cm$ ;

\item
The readout electronics and therefore also cooling
and cabling have to be located very close to the sensors to cope with 
the $96\ns$ bunch crossing time of HERA;

\item
The detector and the frontend electronics need to operate inside a
1.4~T axial magnetic field.
\end{itemize}

The geometrical layout of the MVD consists of a barrel (BMVD) and a
forward (FMVD) section. Here we note that the ZEUS coordinate system
is right-handed with the $Z$-axis along the incoming proton beam
direction, the $X$-axis in the horizontal plane pointing to the inside
of the HERA ring and the $Y$-axis pointing upwards. The barrel
section, centered at the interaction point, is about $63\,$cm long.
The silicon sensors are arranged in three concentric cylindric
layers. The innermost layer is incomplete in the polar angle range of
$60^\circ$ around the positive $X$-axis. The forward section is
composed of four vertical planes extending the angular coverage down
to 7$^\circ$ from the beam line. Access for cables and cooling is
arranged from the rear.

The ZEUS MVD was built in a collaboration of several institutes within the
context of the ZEUS collaboration during 1997 through 2000. It
was installed inside ZEUS in the spring of 2001. In the fall of 2001
the detector recorded first events. From then until the end of HERA
operation in June 2007 the detector has been successfully operated
over a period of almost 6 years. During that time a luminosity of
approximately $430\,{\rm pb^{-1}}$ was collected and more than 600 million
electron-proton scattering events were recorded using the detector.

In the following sections we give a comprehensive description of the
different detector subsystems and we discuss how the requirements set
forth were met taking the constraints into account. We also adress the
conclusions that we have drawn.  The following detector subsystems
will be described in detail:
\begin{itemize}
\item
Silicon Sensors,

\item
Readout Electronics,

\item
Mechanical Construction,

\item
Data Acquisition System,

\item
Slow Control,

\item
Radiation Protection,

\item
Data Quality Monitoring,

\item
Alignment and Tracking.
\end{itemize}
At the end we show some representative physics results obtained with
the detector and we end with a conclusion.

\section{Silicon Sensor Design}
\label{sec:sensors}

The design of the silicon sensors of the ZEUS MVD as well as the
prototype and acceptance tests have been described in detail
in~\cite{nima:2003:505}. Here we just summarize the main features of
the sensors and give some additional information on the ideas behind
the design chosen.

Three different sensor geometries have been designed: square sensors
for the barrel part~(BMVD), and two wedge shaped sensors for the
forward part~(FMVD) of the detector. Table~1 gives 
the relevant geometrical parameters.

\begin{table*}
\begin{center}
\begin{tabular}[t]{|l|c|l|c|c|l|}
\hline type & shape & dimensions & \raggedleft{$p^+$ strip length} & \raggedleft{\# of readout strips} &
active area\\
\hline\hline BMVD &
\begin{minipage}[t]{60pt}\vfill\includegraphics[height=2.5cm]{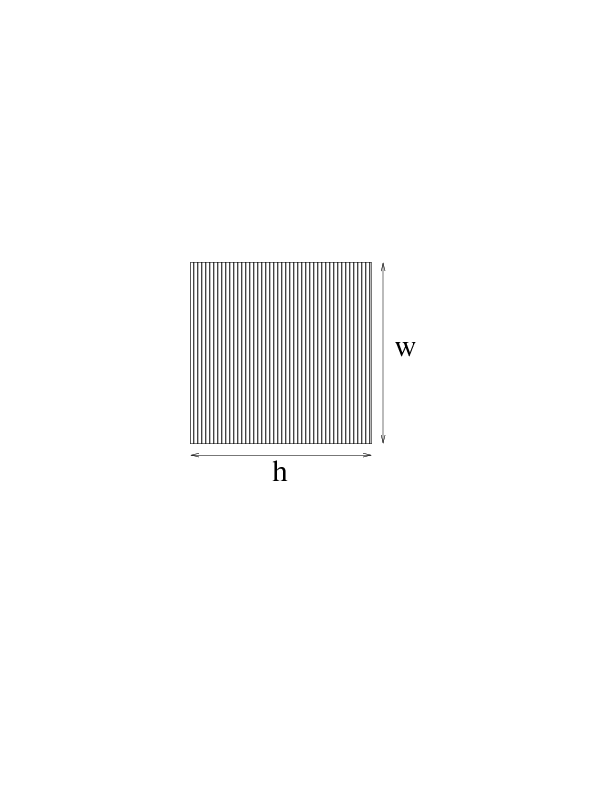}\end{minipage}&
{\parbox[t]{50pt}{h=64.2mm\\ w=64.2mm}} & 62.2mm & 512 & 38.6cm$^2$\\
\hline FMVD1 &
\begin{minipage}[t]{60pt}\vfill\includegraphics[height=2.5cm]{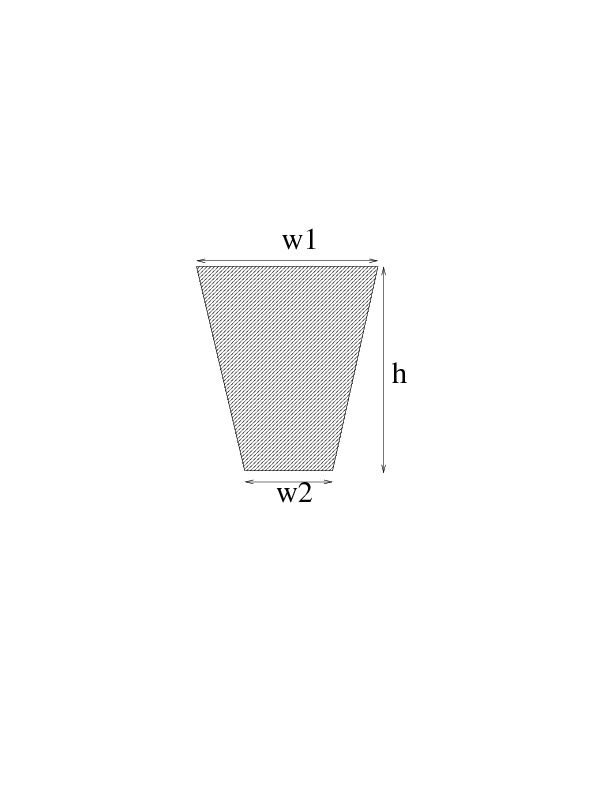}\end{minipage}&
{\parbox[t]{50pt}{h=73.5mm\\ w1=64.3mm\\ w2=30.7mm}} & 5.6 - 73.3mm & 480 & 32.6cm$^2$\\
\hline FMVD2 &
\begin{minipage}[t]{60pt}\vfill\includegraphics[height=2.5cm]{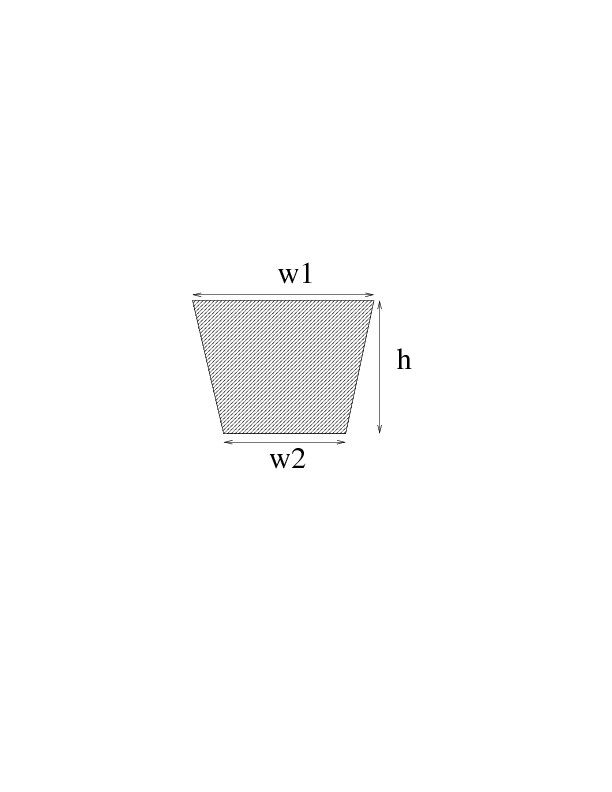}\end{minipage}&
{\parbox[t]{50pt}{h=48.5mm\\ w1=64.1mm\\ w2=42.0mm}} & 5.6 - 47.4mm & 480 & 23.9~cm$^2$\\
\hline
\end{tabular}
\label{tab:sensorgeom} 
\caption{Geometrical parameters of the BMVD and FMVD sensors.}
\end{center}
\end{table*}

\begin{figure*}[hpt]
  \begin{center}
  \includegraphics[height=9cm]{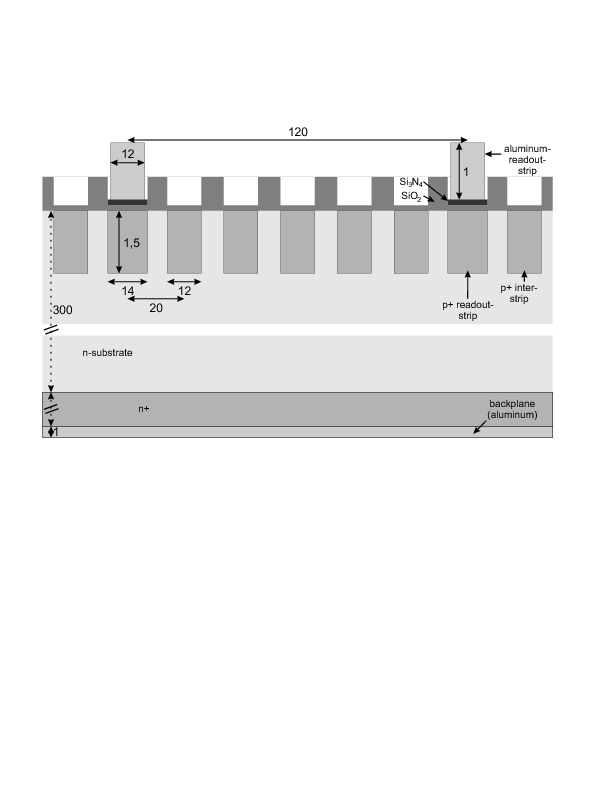}
  \caption{Cross section of the silicon sensors. All dimensions are in
  $\mu m$.}  \label{fig:sensor-xsec} 
  \end{center} 
\end{figure*}

Apart from their geometry, the three sensor types are identical. They
are single-sided strip detectors on high ohmic n-type silicon. The
readout pitch is $120\,{\rm \mu m}$ with 5 intermediate strips for
capacitive charge division. The p$^+$-readout strips are AC-coupled
through a SiO$_2$-Si$_3$N$_4$ double layer to Al-strips, which are
connected via wire bonds either to a Polyimide foil with Cu strips
(Upilex~\cite{upilex}) or a glass fan-out. The p$^+$-readout strips and the
p$^+$-intermediate strips are connected individually via $1.5\,{\rm
M\Omega}$ poly-silicon resistors to the biasing ring. Three
p$^+$-guard rings surround the biasing ring. The innermost guard ring
is connected resistively to ground and collects the current from
defects outside of the sensor's sensitive volume. The two outer guard
rings are left floating. Outside of the last guard ring an n$^+$-line
is placed which allows biasing the detector from the top.  The most
important parameters of the sensors are given in
Table~\ref{tab:sensors}; a sketch of the cross section is shown in
Figure~\ref{fig:sensor-xsec}. The sensor design was guided by the following ideas: 

Analog readout was chosen because of available experience in the group
and the greater ease with which noise, pickup and general performance
could be understood. After exploring different options the
HELIX~128-v3 chip~\cite{helix} was selected for the frontend readout;

Single-sided readout was chosen because of the ease of mechanical
mounting, the larger number of vendors of sensors, costs and, last but
not least, because of a tight time schedule. The compromise of an
additional $0.33\,$\% of a radiation length per two dimensional
coordinate, which increases secondary interactions and multiple
scattering, was considered an acceptable compromise;

The readout pitch of $120\mum$ and need of serial bonding of two
sensors for the BMVD was dictated by the maximum number of readout
channels due to space, the amount of inactive material in the path of
the particles and the cooling requirements. Capacitive charge division
with 5 intermediate strips~\cite{nima:1985:235} with $20\mum$ pitch
does not present a technological challenge for sensor production on
the one hand, on the other hand it results in a nearly linear charge
sharing between the readout strips. Thus the position resolution is
fairly independent of the particle's passage relative to the readout
strip and is essentially determined by the combined sensor and readout
noise. A disadvantage of the capacitive charge division is, that the
measured total pulse height varies by about $30\pc$ as a function of
the particle passage relative to the readout strips. Another feature
of the readout scheme chosen is that the position resolution depends
only weakly on angle up to about 30$^\circ$.  Simulations have shown,
that for single detectors at normal incidence a position resolution of
$8\mum$ is expected for the BMVD-type detectors. The simulation also
shows that the position resolution is directly proportional to the
readout noise. Thus a significant worsening is expected if two
detectors are put in series via a Upilex cable as done for the BMVD;

The performance of single BMVD detectors has been studied in detail in
an electron test beam of $6\gev$ maximum energy at the DESY electron
synchrotron~\cite{thesis:milite:2001,thesis:moritz:2002,mvdtestbeam:0212037}. For
hits with pulse heights below 1.7 times the most probable value (to
avoid degradation due to delta electrons) a resolution of $7.7\pm
0.1\mum$ has been measured for normal incidence. After correcting for
multiple scattering and the resolution of the beam telescope, an
intrinsic resolution of $7.2\pm 0.2\mum$ has been obtained. In
addition the spatial resolutions and pulse height distributions for
incident angles from $0^\circ$ to $70^\circ$ have been
investigated. The results show that the resolution is below $20\mum$
for angles up to $20^\circ$. If a ``head-tail
algorithm''~\cite{thesis:milite:2001,thesis:chiochia:2003} is used, a
resolution below $30\mum$ for angles up to $70^\circ$ is achieved. The
results are compatible with simulations which take into account the
fluctuations of the charge deposition along the particle track, the
readout noise and the electric network of the detector and the readout
amplifiers~\cite{thesis:chiochia:2003,thesis:moritz:2002};

Similar test beam measurements~\cite{thesis:chiochia:2003} have been
performed with BMVD half modules (see section~\ref{sec:mechanics}), in
which two BMVD sensors are connected in series by a Upilex cable. As
expected from the simulations the increase in readout noise due to the
increased capacitance worsens the spatial resolution to about $13\mum$
at normal incidence;

A particular feature of the sensor design is that both bonding and probe pads
are inside the sensitive sensor volume. As the width of these pads is
75 $\mu $m they couple capacitively to the intermediate strips in a
different way than the strips themselves.  Using SPICE simulations, in which the
detector is simulated by a three-dimensional RC network coupled to the
small signal model of the HELIX chip, it has been shown that during
the $50\ns$ shaping time the charge spreads approximately $1\cm$ along the
strips, resulting in a change of the position resolution for tracks
passing the bond pads~\cite{thesis:chiochia:2003,thesis:martens:1999};

Finally the choice of interconnecting sensors with orthogonal strips
(see Section~\ref{sec:mechanics}) was chosen as it meant a significant
simplification of the sub-module construction. Monte Carlo simulations
showed, that such a design neither increases the number of ambiguous
tracks nor causes significant complications for the track
reconstruction.

The sensors have been produced by Hamamatsu Photonics
K.~K.~\cite{hamamatsu}. For details of the measurement of the
technological parameters and the acceptance tests we refer
to~\cite{thesis:martens:1999,nima:2003:505}. We just mention here,
that for the acceptance tests the depletion voltage, the maximum
detector currents and their stability in a long term test have been
measured for every detector. In parallel to production, test
structures were measured to verify technology parameters and the
radiation sensitivity.

Throughout the production the quality of the detectors has been
excellent.
\begin{table}
\begin{tiny}
 \begin{tabular*}{225pt}{|p{90pt}|p{130pt}|}
 \hline
  sensor type & n-type Si, single sided p$^+$ implants\\
  resistivity & $3 < \rho < 8 $k$\Omega$cm\\
  depletion voltage & $40 < V_{dep} < 100 $V\\
  depth active volume  & d = $300~\mu $m\\
  readout strips & Al on p$^+$ strips, AC coupled ($\mathrm{SiO_2 - Si_3N_4}$ double layer)\\
  p$^+$ readout strip pitch & $120~\mu $m\\
  p$^+$ readout strip width & $14~\mu $m\\
 \raggedright{number of interstrips} & 5 \\
  p$^+$ interstrip width & $12~\mu $m\\
  Al readout strip width & $12~\mu $m\\
  backplane & deep n$^+$ layer, aluminized\\
  strip biasing & poly-Si resistor\\
  number of guard rings & 3\\
 \hline
 \end{tabular*}
 \label{tab:sensors}
 \caption{Parameters of the MVD sensors.}
\end{tiny}
\end{table}

\section{Readout Electronics}
\label{sec:electronics}

In addition to the reasons given in the previous section analog
readout also has the added benefit of measuring the specific energy
loss of different particle species in the detector's sensitive
volume. In the following the different components of the readout
electronics, namely the hybrids with the frontend chips, the
patchboxes, the analog link, the ADC and the clock and control
systems are described in more detail.

\subsection{Helix Chips and Hybrids}
\label{subsec:helix}

The HELIX~128-v3 chip has been selected for the frontend readout.  The
HELIX was originally developed by the ASIC Labor Heidelberg for the
HERA-B experiment. It was designed in $0.8\,\mum$ CMOS technology and
was manufactured by AMS~\cite{AMS}. In the following only the general
features of the HELIX chip are described. More details can be found
in~\cite{helix,TRUNK}.

One HELIX chip contains 128 channels, each having its own
charge-sensitive preamplifier, a shaper and a 141-cell analog
pipeline. The bias settings and various other parameters which
determine the shaping time, signal and gain can be adjusted using
programmable DACs. The chip runs at $10.41\mhz$ clock speed
synchronized to the HERA bunch crossing cycle.  Roughly speaking, it
works in the following manner: Once receiving a trigger, the
corresponding 128 analog data signals in pipeline cells are sent out
over a single analog line synchronized with the clock.  At the end of
the transfer, an 8-bit trailer encoding the cell location is
generated. Several chips can be daisy-chained.  In that mode, signals
from several chips are multiplexed and read out over one analog line.
In case of ZEUS, 8 chips are chained together.
\begin{figure}[htb]
	 \begin{center}
    \includegraphics[width=5cm,angle=-90]{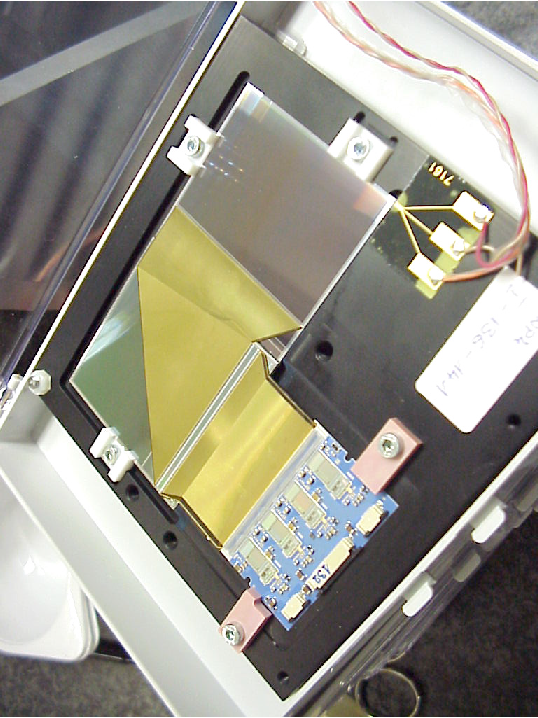}
  \end{center}
  \caption{A ZEUS MVD barrel half module.}
  \label{fig:barrel-module}
\end{figure}
Four chips are hosted on a ceramic hybrid board which
provides bias and supply voltages and routes control and output
signals. The sensors are connected to the hybrids through a flexible
Upilex cable and a glass pitch adaptor. A photograph of a barrel
half module attached to a pair of sensors is shown in
Figure~\ref{fig:barrel-module}.

\subsection{Cabling, Patchboxes and Shielding}
\label{subsec:cabling}

\begin{figure}[htb]
	 \begin{center}
    \includegraphics[width=3.5cm,angle=-90]{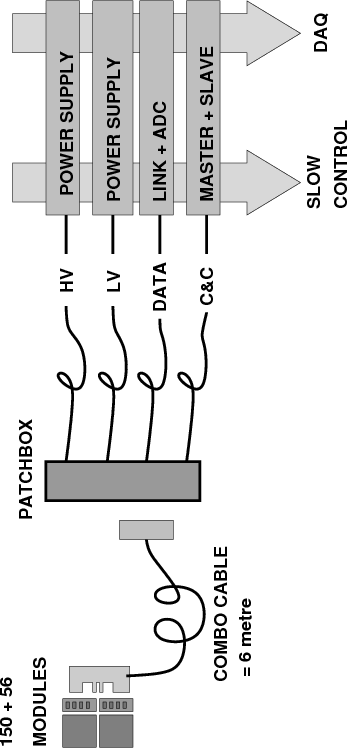}
  \end{center}
  \caption{The readout chain of the ZEUS Microvertex Detector.}
  \label{fig:rochain}
\end{figure}
Figure~\ref{fig:rochain} shows a sketch of the readout chain of the
detector downstream of the detector modules. Supplies and signals as
well as sensor lines for temperature and humidity are routed along the
rear part of the ZEUS beampipe over a distance of ca. $6\,{\rm m}$
using customized ``COMBO'' cables made by Axon~\cite{axon}. In the
barrel the connection between the ceramic hybrids and the COMBO cables
is done using a $500\mum$ pitch flex PCB and miniature ZIF
connectors. This solution was chosen due to the severe space
constraints. In the forward section of the detector these constraints
are less severe and hence standard PCBs were chosen. Just outside the
ZEUS detector on the rear side the COMBO cables are connected to four
patch boxes where bias and supply voltages as well as data and control
cables are split up. The COMBO cables provide double shielding and in
addition groups of COMBO cables are routed inside conductive zipper
tubes for additional protection against electromagnetic
interference. Also, the cables running from the patchboxes to the
readout racks some $20\,{\rm m}$ away are doubly shielded. This
careful shielding allowed to keep the entire analog signal chain
from the sensors to the backend readout system outside the detector
without active elements. This approach was deemed necessary since the
frontend of the detector is not easily accessible and it was planned
to operate it for a period of five years without intervention. In
retrospect this approach has proven very successful.

\subsection{Analog Link}
\label{subsec:analog}

The total number of analog signal lines in the detector amounts to
206. Each HELIX chip also provides an analog
dummy output which follows almost identical circuitry but is not connected to
the sensors and lacks the first preamplification stage. Both outputs
of the HELIX daisy-chain are subtracted and amplified by an analog
link board which feeds it to the ADC module. This analog subtraction
takes care of the dominant contribution of the common noise.  After
the subtraction the signal range is $0\,-\,2\,{\rm V}$, which
corresponds to $0\,-\,10\,{\rm MIP}$.\footnote{1 MIP corresponds to
about 24,000 electrons at the amplifier input.}

\subsection{ADC Module Design}
\label{subsec:adcdesign}

\begin{figure}
  \begin{center}
  \includegraphics[width=.45\textwidth]{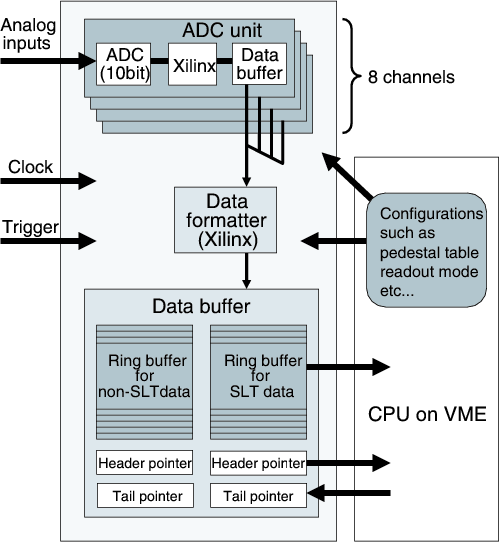}
  \end{center} 
  \caption{Schematic view of the ADC module. The buffer
  for SLT data stores cluster data which is used by the Global Track
  Trigger (GTT) and the buffer for non-SLT data holds raw and/or strip
  data} \label{fig:adcm}
\end{figure}

The ADC module design is described in detail in~\cite{ADC}. Only a
brief description is given here. A schematic view of the ADC module is
shown in Figure~\ref{fig:adcm}.  One ADC module processes 8 analog
input channels.  Each ADC unit (ADCU) handles one analog input line.
It consists of a 10-bit analog-to-digital converter (AD9200) and a
data processor implemented in a programmable gate array (the XC4028EX
of XILINX).  The processed data are stored in FIFOs.  A data
formatter, implemented in another XC4028EX, reads out the FIFOs of all
ADCUs and writes formatted data into data buffers, which are
accessible from the VME bus.

\begin{figure}
  \begin{center}
    \includegraphics[width=7.8cm]{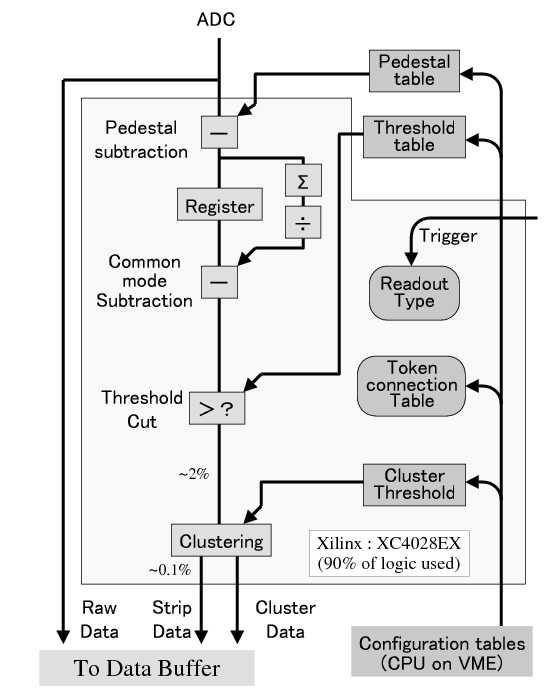}
  \end{center}
  \caption{Block diagram of the Xilinx processor on the ADC unit.}
  \label{fig:process}
\end{figure}

The data processor in the ADCU is similar to that described
in~\cite{adc-scheme}.  The block diagram of the data processor is
shown in Figure~\ref{fig:process}.  First, 10-bit ADC data are
processed for pedestal and common mode subtraction. (At this level,
after the analog subtraction in the Analog Link, the common noise is
typically very small.) Then, strips with charge greater than a seed
threshold are kept for cluster finding.  A second cluster threshold is
set for the total charge of the cluster before further data reduction.
The size of the input raw data volume is reduced to about 2\% after
the threshold cut and to about 0.1\% after the cluster cut.  An ADCU
has 3 FIFOs for the output data: for raw data, for strip data which
consist of strip ID and ADC values after the processing, and for
cluster data which contain the total charge and the location of
the cluster.

The performance of the analog part is monitored with the internal
charge injection of the HELIX which generates a repetitive pattern
equivalent to +2MIP, +1MIP, -1MIP and -2MIP in each channel. The
digital processing part is tested separately with digital test
patterns stored in memory implemented in the module.

\subsection{Clock and Control}
\label{subsec:candc}

\begin{figure}
  \begin{center}
    \includegraphics[width=7.8cm]{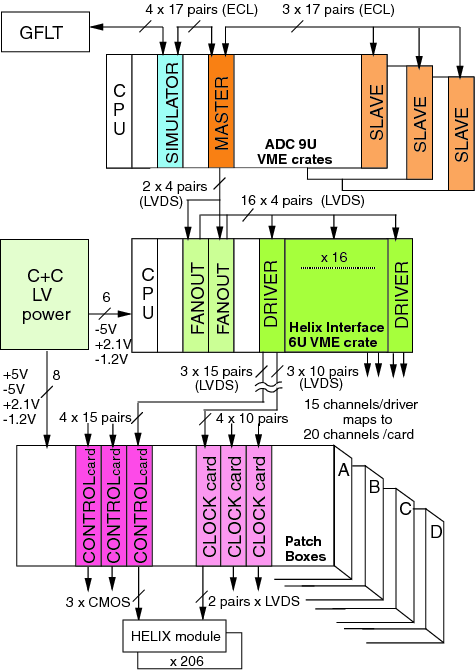}
  \end{center}
  \caption{Block diagram of the C\&C system of the ZEUS MVD.}
  \label{fig:ccscheme}
\end{figure}

The clock and control system (C\&C) consists of three parts: These are
the master and the slave timing controllers, the helix driver modules
and the active electronics inside the patch
boxes. Figure~\ref{fig:ccscheme} shows a schematic of the system. In
the following the major components of the system are briefly
described. Many more details may be found in~\cite{ccref}.

The master and slave timing modules handle the fanout of the trigger
and clock signals from the ZEUS Global First Level Trigger (GFLT).
The main functions of the C\&C master are to receive the standard GFLT
information and to pass it on to the C\&C slaves, decode the trigger
information in order to produce and send the relevant signals to the
HELIX fanout modules together with a correctly timed $96\,$ns clock
and to handle the \texttt{ERROR} and \texttt{BUSY} signals from the
ADCs. Additionally, the C\&C master is able to operate in a
stand-alone mode for test purposes, generating its own $96\,$ns clock.
The C\&C slaves receive the GFLT signals from the C\&C master and
distribute them to all the ADCs within their crates and they receive
back the \texttt{BUSY} and the \texttt{ERROR} lines from the
ADCs. They also distribute the suitably timed $96\,$ns clock
synchronously to all the ADCs in their crates, via a specially
wired-in set of 16 twisted pairs of identical length, to front panel
connectors. The C\&C slaves also allow a continuous VME read access to
a register with the ADC-produced signals. Additionally, the C\&C
slaves can operate with the C\&C master as a standalone system for
test purposes.

16 Helix Driver Modules are located in a separate VME crate. Each
Helix Driver module generates the required controls signals for 15
Helix channels from the clock and control commands received from the
C\&C master via two Helix Fanout modules. Programming information for
the Helix modules in the detector is sent along the VME back-plane to
the Helix Drivers, then after some processing to the Patchboxes and on
to the detector. The five C\&C signals are \texttt{CLOCK},
\texttt{TRIGGER} (this doubles as the Serial Data line during
downloading), \texttt{TEST PULSE}, \texttt{LOAD} and 
$\overline{\texttt{RESET}}$.

Each Helix Driver board carries two Programmable Logic Devices (PLDs),
one of which contains the VME interface electronics, and the other,
the on-board processing logic, status registers, etc. The incoming
signals are treated as follows: The clock signal is fanned to 15
clocks, one for each channel, and each with its own programmable delay
unit. The purpose of these is to remove timing skews at the
detector. The trigger signal is counted, then OR-ed with the serial
data line used for downloading. The combined signal is then sent to
all 15 channels.  The test pulse is expanded to 15 channels, which are
compared with a mask register, then the unmasked ones output to
selected channels.  The $\overline{\texttt{RESET}}$ is OR-ed with an internally
generated $\overline{\texttt{RESET}}$ signal used during downloading. The combined
signal is sent to all channels.

The C\&C boards in the patchboxes both have active components. The
five signals per channel arriving from the Helix Drivers are slightly
degraded at the end of the 20 meter cables, and in the case of the
clock boards, the signals are re-generated by a receiver-transmitter
combination (LVDS to $3.3\,{\rm V}$ CMOS/LVDS) before being sent down the
micro-cable twisted pairs to the Helix Modules. This precaution is
necessary due to the relatively high attenuation factor of the
micro-cables.

\section{Mechanical Construction}
\label{sec:mechanics}

\subsection {Overall layout}

\begin{figure*}[htbp!]
\begin{center}
\includegraphics[width=15cm]{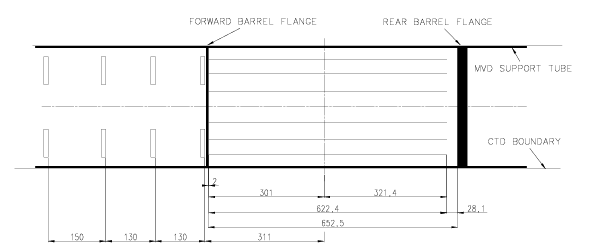}
\end{center}
\caption{Layout of the MVD along the beamaxis; protons go from right to left.} 
\label{layout}
\end{figure*}

The mechanical layout of the MVD consists of a barrel and a forward
section. A sketch of the cross section along the beam line is shown in
Figure~\ref{layout}.We recollect that the ZEUS coordinate system is
right-handed with the $Z$-axis along the incoming proton beam
direction, the $X$-axis in the horizontal plane pointing to the
inside of the HERA ring and the $Y$-axis pointing upwards.

\begin{figure*}[htbp!]
\begin{center}
\includegraphics[width=12cm]{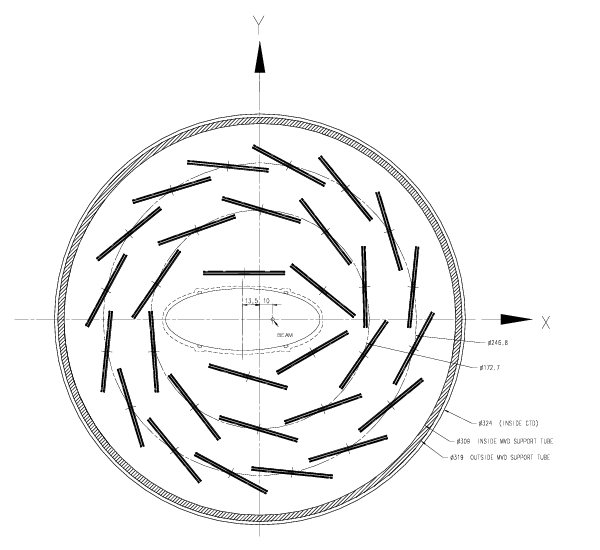}
\end{center}
\caption{Cross section of the layout of the silicon sensors in the barrel MVD. The direction in which protons travel points out.}
\label{xybarrel}
\end{figure*}

The barrel section, centered at the interaction point, is about
$63\,$cm long.  The silicon sensors are arranged in three concentric
cylindric layers. As shown in Figure~\ref{xybarrel}, approximately
25\% of the azimutal angle is covered by only two layers due to
limited space.  The polar angular coverage for tracks with three hits
ranges from 30$^0$ to 150$^0$. The modules in the forward section are
arranged in four vertical planes extending the angular coverage down
to 7$^\circ$ from the beam line. Exits for cables and cooling is
arranged from the rear.  Many details, design drawings and pictures
can be found in~\cite{mvdweb}.

\subsubsection {Support tube}

The MVD had to fit inside the inner cylinder of the Central Tracking
Detector (CTD), a cylinder with a diameter of 324~mm. The only
available support points are located at the forward and the rear side
of the CTD, a span of $2\,{\rm m}$. To make efficient use of the
available space it was decided to install the MVD and the beampipe
together; the support for the MVD has then to be made in two
halves. The two half cylinders are joined together, providing a stiff
support for the whole assembly. During and after installation the
weight of the beampipe has to be taken by external supports.

\begin{figure*}[htbp]
\begin{center}
\includegraphics[width=12cm]{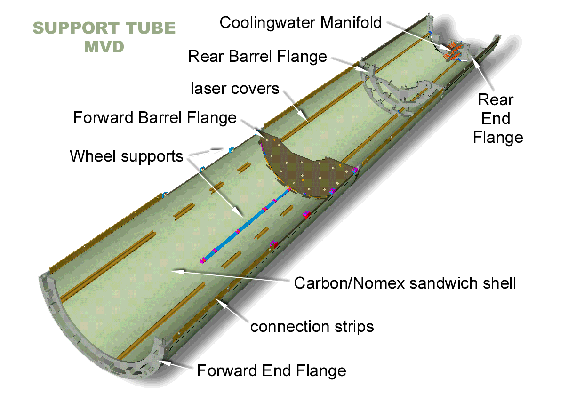}
\end{center}
\caption{Elements forming a half support tube.}
\label{supporttube}
\end{figure*}

The support tube is an assembly of light weight half cylinders
connected to each other via flanges.  A $4\mm$ thick honeycomb layer
is glued between two carbonfiber ($0.4\mm$ thick) sheets and preformed
on a cylindrical mold in an autoclave. During this process the inside
and outside of the tube is covered with a $25\mum$ thick aluminium
layer for electric shielding. Figure~\ref{supporttube} shows a sketch
of one half of the support tube.

\subsection {Barrel modules}

Details on the silicon sensors, their
characteristics and performance are described in
Section~\ref{sec:sensors}.

\begin{figure}[htbp!]
\begin{center}
\includegraphics[height=15cm]{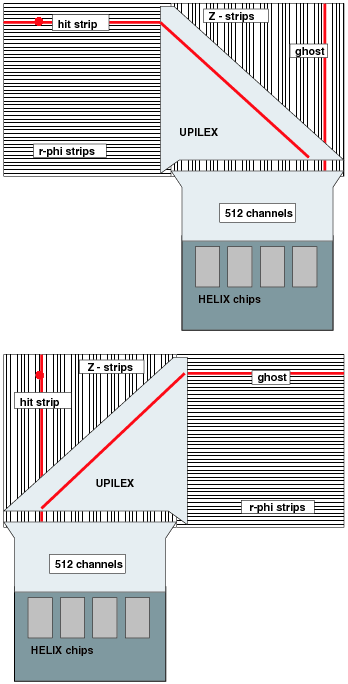}         
\end{center}
\caption{Two silicon sensors are assembled into a half module; the two
half modules are mounted on top of each other and form a full module.}
\label{module}
\end{figure}

Two sensors are glued together and one sensor is electrically
connected to the other as shown in Figure~\ref{module}, forming a {\em
half module}.  For various reasons we chose to orient the readout
strips of the two sensors within one half module perpendicular to each
other as shown in Figure~\ref{module}. A
Cirlex~\cite{mvd-mechanics:cirlex} strip ($5.8\times 65.44\mm$,
$0.4\mm$ thick) glued at the edge in between the sensors, forms
the mechanical connection between the two. The use of this material
ensures a good insulation of the HV side of one sensor from the
ground plane of the other one.

While the sensors overlap by $4.8\,{\rm mm}$, the sensitive areas only
overlap by $2\mm$. Cirlex feet, with a cross section of a few square
mm and varying in thickness from 0.8 to $2.4\mm$, are used to
connect the sensors to the support structure.

The electrical connection between the two sensors is made via copper
traces on a $50\mum$ thick Upilex foil. The foil is precisely aligned,
glued on the sensors and subsequently wire bonded to the strips. The
connection of the sensor assembly with the frontend readout is also
made via a Upilex foil, glued at one side on a sensor and at the other
side on the hybrid as indicated in Figure
\ref{module}.

A surface of $123.68\times 64.24\,{\rm mm^2}$ forms one readout cell
of 512 channels and is called a {\it half module}. A mirror image of
this half module is also shown in Figure~\ref{module}; both are
mounted on top of each other, separated by the spacers mentioned above
forming a {\it full module} with 1024 readout channels.

Two precision markers, glued to each half module, provide reference
points for the alignment on the support structure.

\subsubsection {Barrel module assembly}

The area covered by the silicon in the barrel section is $622.4\,{\rm
mm}$ long and is formed by two cylinders with radii of approximately
125 and $90\,{\rm mm}$.  The coverage is realized with segments of
$\sim 64.2\times 622\,{\rm mm^2}$ arranged like roof tiles, such that
there are no gaps in azimuth angle between adjacent segments for the
outer two cylinders. A third layer close to the beampipe provides a
coverage of about $75\,{\rm \%}$ in azimuth angle.
Figure~\ref{xybarrel} shows a cross section of the silicon sensors in
the barrel section.

Five full modules are fixed side by side, with $1\mm$ gaps in between,
on a support frame and form a ``ladder''.  The modules are glued to
the ladder with the spacers mentioned earlier.

The frontend electronics, cabling and cooling are attached to the
support frame that holds the modules: a ladder.  In order to conserve
the high resolution properties of the silicon sensors a support
structure that bends less than $25\mum$ when all components are
mounted was required.

\begin{figure}
\begin{center}
\includegraphics[width=0.5\textwidth]{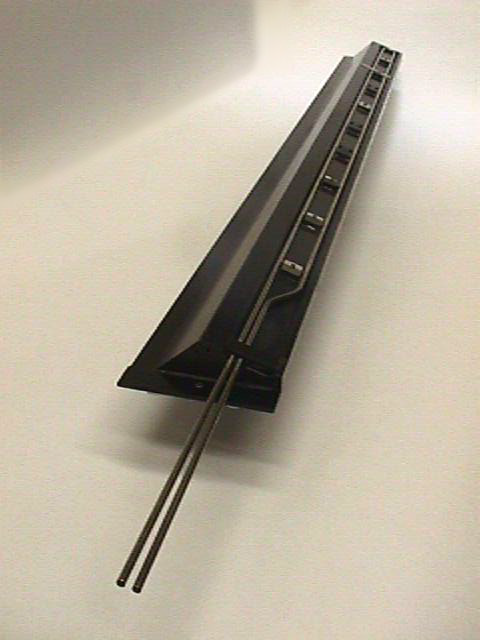}
\end{center}
\caption{Bare support structure for barrel modules}
\label{ladder1}
\end{figure}

\begin{figure}
\begin{center}
\includegraphics[width=0.5\textwidth]{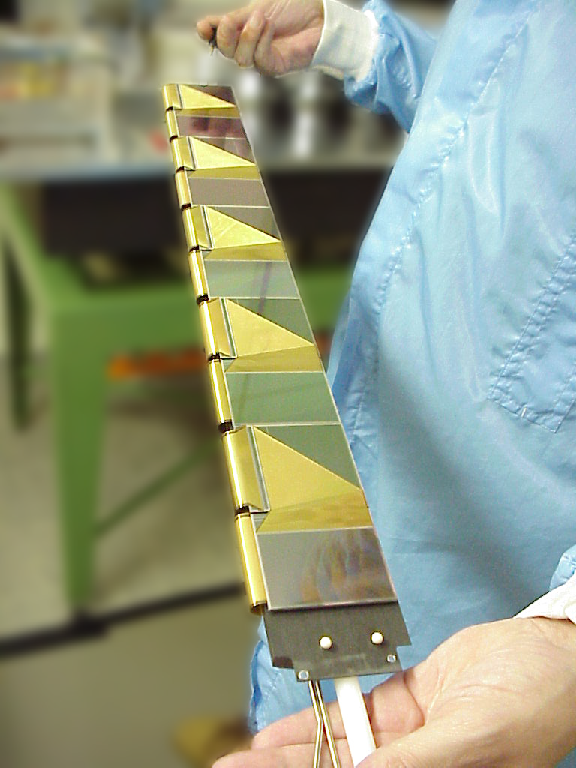}
\end{center}
\caption{Silicon sensors mounted on ladder}
\label{ladder2}
\end{figure}

\begin{figure}
\begin{center}
\includegraphics[width=0.5\textwidth]{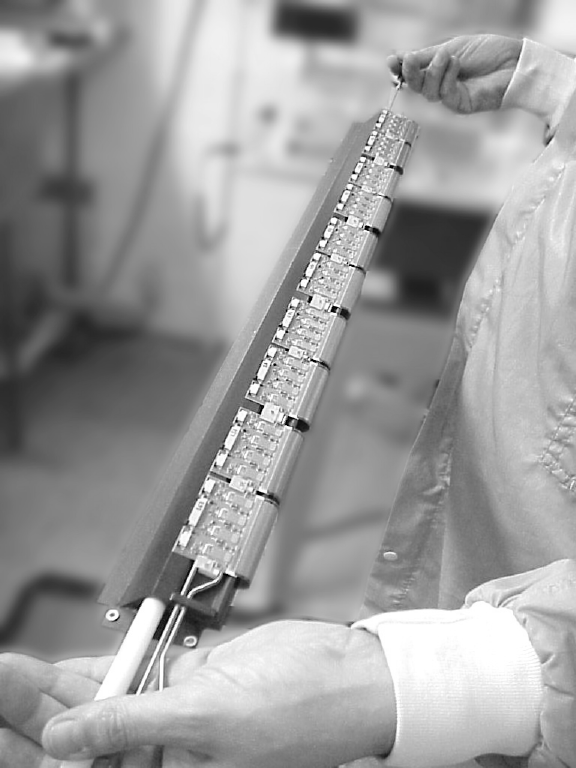}
\end{center}
\caption{Hybrids mounted on ladder}
\label{ladder3}
\end{figure}

The total weight to be carried by one ladder, silicon plus
electronics, cooling and cabling is $\sim220\,{\rm g}$.  The cross
section of the support structure is triangular to accommodate a
support for the hybrids and cooling, providing at the same time the
necessary stiffness.  Carbonfiber material (five layers of fiber with
a total thickness of $0.4\,{\rm mm}$) is used to realize a strong and
lightweight construction.  Strips of carbonfiber material with
variable width and approximately $65\,{\rm cm}$ length are cut from
the sheets and glued together with the help of a mold to form a
ladder.  No deformation has been observed as a function of humidity or
temperature changes. Figure~\ref{ladder1} shows the bare ladder with
water cooling pipe (details about the water cooling are described in
section~\ref{sec:sc}), a stainless steel tube with 0.1mm wall
thickness.  Figure~\ref{ladder2} shows the ladder with the sensors
mounted, while Figure~\ref{ladder3} shows the mounting of the frontend
electronics; the flexible Upilex connection between sensor and hybrid
allows fixation of the hybrid on the side of the ladder on top of the
cooling pipes. Table \ref{inventory-barrel} gives an overview of the
parts forming the barrel detector.

\begin{table*}[tbp]
\vspace{5mm}
\begin{center}
\begin{tabular}{|l|r|r|r|r|} \hline
  & inner cylinder & medium cylinder & outer cylinder& Totals \\
\hline
 radius(mm) &  & 90 & 125 &    \\
\hline
 number of ladders & 4 & 10 & 16 & 30 \\
\hline
 number of silicon sensors & 80  & 200 & 320  & 600 \\
\hline
 number of readout cells & 40 & 100 & 160 & 300 \\
\hline
 number of readout channels & 20480 & 51200 & 81920 & 153600 \\
\hline
 number of readout chips & 160 & 400 & 640 & 1200 \\
 (128 channels/chip)     & & & &     \\
\hline
\end{tabular}
\end{center}
\vspace{5mm}
\caption{Statistics of the components of the barrel detector.}
\label{inventory-barrel}
\end{table*}

\subsection {Forward modules}

\begin{figure}[htbp!]
\begin{center}
\includegraphics[width=0.5\textwidth]{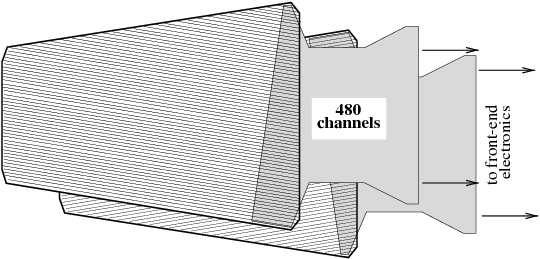}
\end{center}
\caption{Layout of the wheel sensors with the Upilex connections to the hybrids;
the Upilex connection is longer for those sensors which have the
hybrid mounted at the outside of the wheel.}
\label{wheel-layout}
\end{figure}

\begin{figure}[htbp!]
\begin{center}
\includegraphics[width=0.5\textwidth]{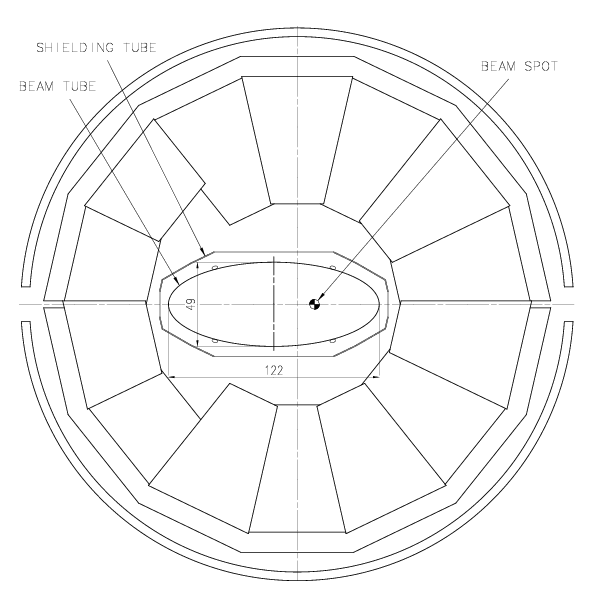}
\end{center}
\caption{Wheel geometry with beampipe and shielding tube, shown inside the
overall support tube}
\label{wheel-geometry}
\end{figure}

Apart from their shape the silicon sensors covering the forward region
are similar to the barrel sensors as described in
Section~\ref{sec:sensors}.  Fourteen wedge-shaped sensors cover a
circular plane around the beampipe. Figure~\ref{wheel-layout} shows
the geometry of one sensor, together with the Upilex foil guiding the
signals to the hybrid.  The strips run parallel to one tilted side;
each sensor has a total of 480 readout strips with $120\,{\rm \mu m}$
readout pitch. Figure~\ref{wheel-geometry} shows the geometric layout
of the sensors for one wheel.

\begin{figure*}[htbp!]
\begin{center}
\includegraphics[width=13cm]{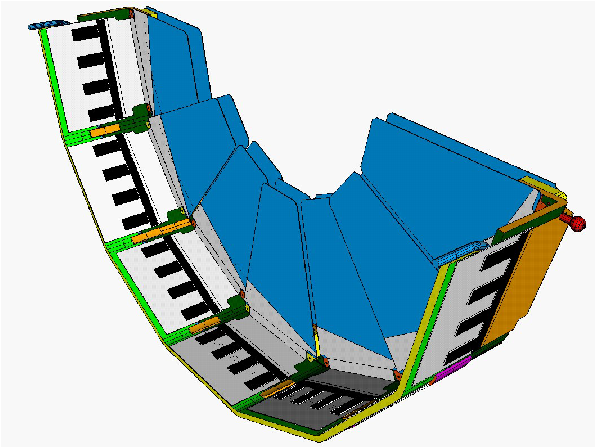}
\end{center}
\caption{Sketch of the layout of a half wheel with sensors, hybrids and
support structure}
\label{wheel-sketch}
\end{figure*}

The FMVD consists of four planes perpendicular to the beam axis; each
plane has two sensor layers.  Adjacent sensors are displaced
perpendicular to the beam line by $3\mm$ and overlap in azimuth by
$4\mm$ ($2\mm$ in sensitive area).  The two layers are mounted back to
back on a support structure, separated by approximately $8\mm$ in
Z-direction. Figure~\ref{wheel-sketch} shows a sketch of the layout of
one half of a wheel. The sensors are mounted on an L-shaped carbon fiber half
ring, with seven straight sections. Aluminum inserts in the short
'leg' of the ring guarantee precision holes for the positioning of the
sensors. These are put back to back, such that the angle between the
strips is $2\times 13^\circ$.  The long side of the L-shaped
support provides housing for the cooling pipes and supports the
hybrids. The half wheels are supported at three points inside the main
support tube.

Tab.~\ref{wheel-inv} summarizes the main parameters of the wheel
geometry and the statistics of the individual parts.

\begin{table}[tbp]
\vspace{5mm}
\begin{center}
\begin{tabular}{|l|r|} \hline
 outer radius(mm) & $\sim$132mm \\
\hline
 inner radius(mm) & $\sim$60mm\\
\hline
 wedge angle    & $\sim$13$^\circ$  \\
\hline
 readout strip pitch    & 120 $\mu$m \\
\hline
 number of channels/sensor & 480  \\
\hline
 number of sensors/wheel& 2 x 14 \\
\hline
 number of wheels & 4 \\
\hline
 total number of sensors& 112 \\
\hline
 total number of readout channels & 53760 \\
\hline
 number of readout chips & 448 \\
 (128 channels/chip)     &     \\
\hline
\end{tabular}
\end{center}
\vspace{5mm}
\caption{Statistics on the components for the forward detector.}
\label{wheel-inv}
\end{table}

\begin{figure}
\begin{center}
\includegraphics[width=0.5\textwidth]{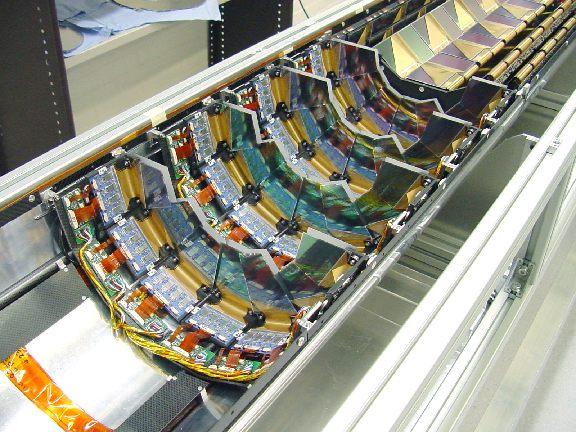}
\end{center}
\caption{Half wheels installed in the overall support structure}
\label{wheel-true}
\end{figure}

Figure~\ref{wheel-true} shows four half wheels mounted in the support
structure.  

\subsection{Material distribution}

\begin{figure}
\begin{center}
\includegraphics[width=0.25\textwidth,angle=-90]{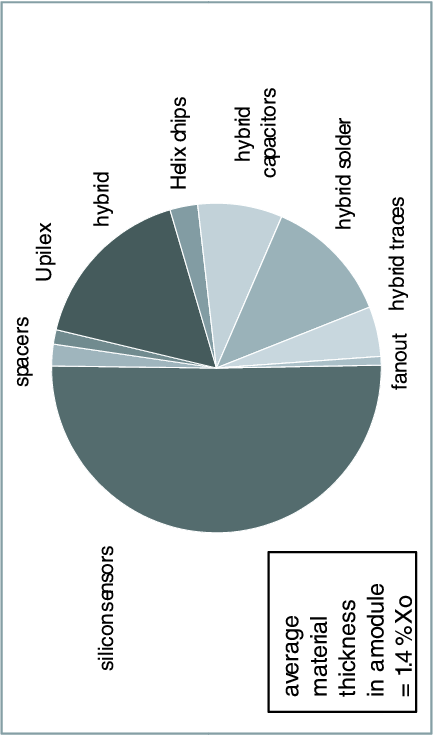}
\end{center}
\caption{Material distribution within a module}
\label{matmodule}
\end{figure}

\begin{figure}
\begin{center}
\includegraphics[width=0.25\textwidth,angle=-90]{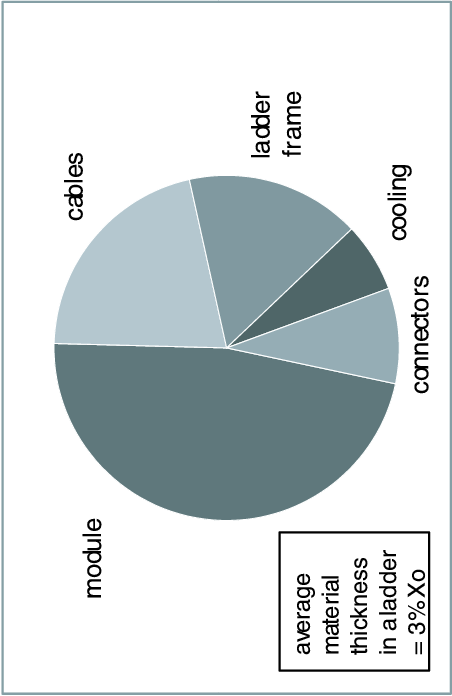}
\end{center}
\caption{Material distribution within a ladder}
\label{matladder}
\end{figure}

The material distribution has been calculated for individual modules
and complete ladders and is summarized in Figs.~\ref{matmodule} and
\ref{matladder}.  If all the material which makes up a module is
equally distributed over its surface, the total thickness represents
$1.4\,{\rm \%}$ of a radiation length; for a ladder this amounts to
$3.0\,{\rm \%}$ (including the modules).

\section{Data Acquisition}
\label{sec:daq}

The architecture of the MVD data acquisition system~\cite{mvddaq} has
been strongly influenced by its contribution to the ZEUS trigger.
The ZEUS data acquisition system is based on a three level trigger.
Because of the HERA bunch crossing rate of $10.4\mhz$, i.e. $96\ns$
between consecutive bunch collisions, the experiment uses a pipelined
readout design. The {\it Global First Level Trigger} (GFLT), based on
a reduced set of information from the detector components, is issued
synchronously after 46 bunch crossings and reduces the trigger rate to
$\le 600\hz$.  Detector data, stored in deadtime-free analog or
digital pipelines, is subsequently digitized, buffered and used by the
{\it Global Second Level Trigger} (GSLT). The GSLT lowers the trigger
rate to $\leq70\hz$ with a typical latency of $10\ms$.  For accepted
events the complete detector information is read out, merged with
data from other detector components by the {\it Event Builder} (EVB) and
sent to the {\it Third Level Trigger} (TLT) computer farm where event
reconstruction and final online selection are performed.  In normal
conditions the total deadtime caused by the detector and trigger
components is about $1\pc$. More details on the ZEUS trigger and data
acquisition system can be found in~\cite{zeusdaq}.

\subsection{The Global Tracking Trigger}

Simulation studies on data multiplicity and background had shown that
the MVD data alone, providing up to 3 planes of information per track,
would not be sufficient for unambiguous tracking and efficient rate
reduction. Combining the MVD information with the surrounding tracking
detectors would instead allow a far better rate reduction and tracking
efficiency.  These considerations have lead to the design within the
MVD data acquisition system of the Global Track Trigger
(GTT)~\cite{gtt}, a distributed computing environment processing data
from the MVD, the Central Tracking Detector (CTD) and the newly
installed forward Straw Tube Tracker (STT).

For the computing infrastructure, the preferred choice has been to use
whenever possible, commercial off the shelf equipment easily
upgradable and maintainable. Investigations on the performance
achievable in terms of data throughput, process latency and
performance suggested a solution based on embedded VME computers
running a realtime operating system for the data readout and a farm of
standard PCs connected via a Fast/Gigabit Ethernet network for trigger
and data processing.

For the readout platform Motorola PowerPC single board CPUs and the LynxOS
operating system were chosen due to the following requirements:
\begin{itemize}
\item Realtime Unix-like operating system;
\item Network boot and diskless operation;
\item Reliable Fast Ethernet network operation;
\item Fast and flexible DMA and Interrupt capable VMEbus bridge. 
\end{itemize}

28 MVD ADC modules are organized in three 9U VME crates for BMVD
upper, BMVD lower and FMVD respectively. These crates receive clock
and trigger information via the Clock and Control Slave Module located
in the same crate. This system is described in more detail in
Section~\ref{subsec:candc}. A short overview is repeated
here. Busy and error conditions are set by the ADC modules and the
VME processor and sent through the Slave Module to the Master Module.
The Master module is directly connected to the GFLT, and transmits
clock and trigger information both to the readout and to the MVD front
end and can be operated via VME to allow standalone MVD running. Data
gathering from the CTD and STT systems, both of which are based on a
network of transputers\footnote {INMOS Transputers, were an advanced
technological development in the late 80's when the ZEUS experiment
was designed.  Provided with a 32 bit processing unit, on board
memory, four 20~MHz serial links for processor interconnection and a
high level parallel programming language (OCCAM), transputers were
ideal for highly distributed parallel processing and data transfer.}
hosted in several VMEbus crates, is done using the same VMEbus
computers~\cite{ctdint}.

On GFLT accept, event data from MVD, CTD and STT are sent to one GTT
processor where the required trigger calculations are performed. The
GTT result is forwarded to another PowerPC VME computer which
re-orders the events according to the GFLT number and transfers the
result to the GSLT.  Since the GSLT is also based on a transputer
network a similar solution as for the CTD and the STT was adopted. No
special hardware is required to receive the GSLT trigger decision as
this is transferred back via TCP/IP.  An interface process forwards it
to the MVD readout CPUs and the GTT processors. On GSLT accept the
full MVD and GTT data are sent to the EVB interface.  It merges and
formats these data into the final format.  Performance and data
quality monitoring is also performed on this system. The specific
hardware components of the MVD DAQ and the GTT systems are summarized
in Table~\ref{tab.hw}.

\begin{table*}[htb]
\begin{tiny}
\begin{tabular}{@{}rll}
\hline
Number & Item & Purpose \\
\hline
\multicolumn {2} {l} {Detector readout and data gathering:}&\\
3 & Motorola MVME2400 450MHz 64MB	& MVD Readout ($3\times$9U ADC VMEbus crates)\\
3 & Motorola MVME2400 450MHz 64MB	& CTD axial, CTD z by timing, STT data\\
\multicolumn {2} {l}{GTT processing:}&\\
10& SuperServer 6014P Dual 3GHz 1GB	& GTT Algorithm Processing\\
1 & DELL PowerEdge 4400 Dual 1GHz 256MB	& GTT Server+Credit/GSLT Decision\\
1 & Motorola MVME2400 450MHz 64MB	& GTT to GSLT Trigger Result Interface\\
1 & DELL PowerEdge 4400 Dual 1GHz 256MB	& Data collection and EVB interface \\
\multicolumn {2} {l}{Supporting network and processing infrastructure:}&\\
1 & DELL PowerEdge 6450 Quad 700MHz 1GB	& NFS File Server, Process Hub and Run Control\\
1 & DELL PowerEdge 4400 Dual 1GHz 256MB	& EVB interface and online monitoring\\
1 & Motorola MVME2700 367MHz 64MB	& LynxOS Boot Server and Development Node\\
1 & Motorola MVME2700 367MHz 64MB	& HELIX frontend programming and monitoring\\
2 & Intel Express 480T Fast/Giga-16Port Cu Switch & Network Connections\\
2 & CISCO XXX switches & Network Connections\\
\hline
\end{tabular}\\
\caption{MVD Data Acquisition and Global Track Trigger 
Computing and Network Resources.}
\label{tab.hw}
\end{tiny}
\end{table*}

\subsection{The VMEbus Readout Software}

In order to fully exploit the VMEbus access capabilities of the
on-board Tundra universe II VME bridge, a dedicated software package
for VMEbus access on Motorola PPC boards running LynxOS was
developed~\cite{uvmelib}.  The package consists of a library layered
on an enhanced driver with respect to the default version distributed
by LynxOS, providing flexible VME memory mapping, DMA transfer and
queueing, VME interrupt and process synchronization control in a
multi-user environment. User programs perform all VME and related
operations using the library without any direct connection to the
driver.

\begin{figure*}[htb]
{\centerline{\includegraphics[width=6cm,angle=-90]{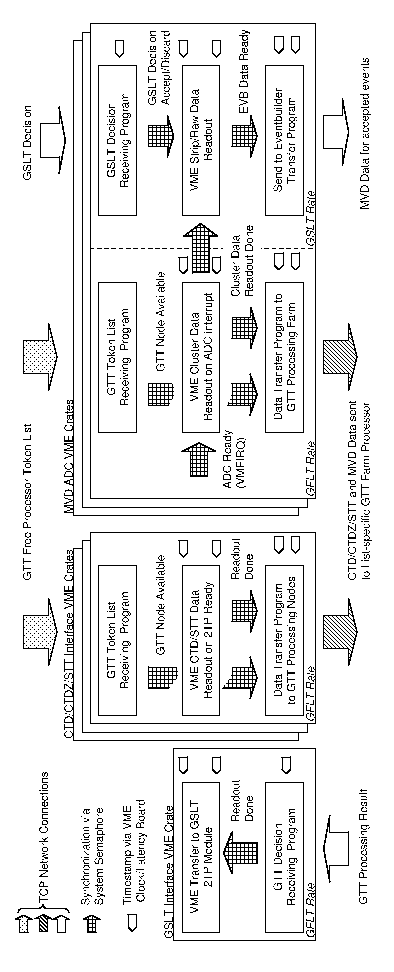}}}
\caption{Block Diagram of the DAQ VME software implementation.}
\label{fig:vmesoft}
\end{figure*}

Within the library, basic data structures to describe Shared Memory
Segments opened both on the internal contiguous memory and on the
VMEbus are available for standard and DMA-controlled read-write
cycles. An additional interface is provided to support process
synchronization by waiting on, or setting system semaphores.  It is
possible to connect VME interrupt handling to system semaphores,
identically treating synchronization to hardware interrupts, DMA
cycles or software signals.  The ability to uniquely identify segments
by ID or name allows many processes to connect to already existing
mapped regions (up to eight in the VME space) without overloading the
system. These features have allowed a modular design in a complex DAQ
environment. A diagram of the main processes running on the VME
computers when taking data is shown in Figure~\ref{fig:vmesoft}.

On the ADC systems two {\it software pipelines} running at FLT and SLT
rates respectively, exist. At FLT rate the readout program running at
lower priority, is woken up on VME interrupt as soon as data are ready
on the ADC boards. After data has been transferred via DMA to internal
memory, the network tasks, synchronized by a semaphore and running at a
higher priority, will send the data to the GTT farm. A similar
software pipeline is available also on the CTD and STT interfaces for
the data read out at GFLT rate.

\subsection{The GTT Environment Process}

The GTT, as any of the components participating in the ZEUS second
level trigger, has to process events at an average rate $\le 600\hz$
with a mean latency, including all the data transfers, of less than 
$10\ms$ and small tails due to busy events or performance fluctuations.

The {\it GTT environment process} is a multi-threaded program with one
thread per input data source ($3\times$MVD, $2\times$CTD,
$1\times$STT), one thread per trigger algorithm and a time limit
thread. Typically one environment runs on each of the dual
CPU farm PCs.  The scheduling of dispatching data to environments is
organized using a synchronized ordered list rather than a round-robin
scheme. The decision for that mechanism was based on simulation
studies.

For development and performance tests a {\it playback} capability has
been provided: This allows one to feed simulation or previously saved
events into the component VME interfaces and into the GTT trigger
chain.

All network data transfers are performed using standard TCP/IP
protocol.  To cope with the different platforms involved (PowerPC for
the VME readout and standard PC for the DAQ and GTT computing nodes)
the communication and synchronization is done via short XDR encoded
messages while detector data is sent with no additional overhead. To
monitor the readout latencies and network transfer times precisely, a
general purpose 6U VME board~\cite{vmeallp} providing a {\it
latency/clock} register, has been developed and installed in all VME
crates and connected to a common $16\,{\rm \mu s}$ clock bus. Thus
absolute timestamps and latency measurements are available for every
event and stored in the data at several points of the data acquisition
phase. This allows a detailed monitoring of the system performance.
During normal running a mean latency smaller than $10\ms$ with steep 
tail for the whole GTT system, compatible with the initial requirements 
is observed. 

\subsection{Computing Environment and Software}

As described previously the DAQ computing environment consists of VME
board computers running LynxOS and Intel PCs running Linux
interconnected via point-to-point Fast or Gigabit Ethernet.  The VME
systems are diskless and are booted over the network from a single,
dedicated, VME system acting as a boot server. All Linux systems are
booted from local disks. In order to simplify code usage a single DAQ
file system, containing executable directories etc., is mounted via
NFS by all participating hosts.

Standard C programming has been used throughout the DAQ and GTT
systems.  A number of ROOT-based C++ GUIs have been developed to
control the DAQ system in stand-alone mode, to view online or archived
histograms, and to manually control the detector slow control system,
etc.

All non-readout or trigger data messages are transmitted through a
{\it hub} process and not directly between processes. The hub is a
multi-threaded process which accepts connections at a known address
and enforces a simple protocol, using XDR.

\subsection{Run and Process Control}

The MVD run control can run in either standalone mode or as part of
the ZEUS run control system.  In both cases process control, stopping
and starting tasks, is facilitated by daemon processes started at boot
time. These advertise themselves to the run control system and
identify, by name, processes they are capable of running. Run
configuration requests from ZEUS or the standalone run control specify
a {\it Runtype} definition file naming all the processes to be started
or stopped, the parameters they require, their required exit status,
etc.  The C preprocessor (cpp) is then invoked to expand all of the
definitions contained in the file (resolving required processes, their
supporting daemon, startup parameters, etc.). The file produced
corresponds to a sequential list of process control commands required
to perform all the transitions specified. The run control system can
then sequence all the steps required for each transition request
received. Transition requests fail if a process fails to reach or does
not remain in its final state.

\emph{}\section{Slow Control}
\label{sec:sc}

Power supplies and other devices required for operating the MVD are
controlled and monitored by the slow control system. The field bus used
to communicate with devices is CANbus, which implements a serial data
transfer protocol with differential signal levels. All devices used
are located at the MVD rack platform and the maximum bus length is
$\sim$10 m. Three ESD CAN-PCI/331~\cite{ESD} dual CANbus adapters
located in two Linux PCs provide the seven buses used to individually
connect all devices. By connecting devices from the same manufacturer
to a bus, baudrate and CANbus addressing incompatibilities were
avoided. The slow control devices are described below followed by a
brief description of the software used.

\subsection{Silicon detector bias voltage}

A commercial high voltage system~\cite{ISEG} is used to provide bias
voltage to the silicon detectors. Four 6U EURO crates (ISEG ECH 238L
UPS), with eight voltage boards (ISEG EHQ F0025p) of 16 outputs each
provide the 412 channels that are required to supply each silicon half
module independently.  The output voltage range is 0 to $+200\,$V,
adjustable in steps of $5\,$mV. The maximum current is $0.5\,$mA.  The
nominal detector bias voltage and trip settings were $60\,$V
and $50\,{\rm \mu A}$, respectively.  The measurement resolutions of
$10\,$nA and $5\,$mV allow accurate detector I-V curves to be
generated for use in monitoring radiation damage.  Two CANbus buses
are used; one to connect the four crate controllers, and another to
connect the voltage boards via the crate backplanes.

An additional non-crate mounted voltage board supplies the MVD
radiation monitor diodes. This device is connected separately to
CANbus.

The ability to control the properties of single and groups of output channels
simultaneously was used extensively during commissioning of the detector. During
data taking this flexibility was not required. 

\subsection{Helix hybrid low voltage}

The frontend hybrid low voltage system is a custom development derived
from the system used by the ZEUS Leading Proton Spectrometer
detector~\cite{LPS}. Six 6U EURO crates, each with 8 voltage boards of
10 outputs provide the 206 channels required to supply $\pm 2\,$V to
each hybrid board with a maximum current of $1\,$A. CANbus is
interfaced to the internal ${\rm I^2C}$ bus at the crate controller
board.

An additional crate, whose boards are modified to output $\pm 5\,$V
and $\pm2.1\,$V, is used to supply voltage to the Helix clock and
control interface electronics.  The CANbus of this crate is connected
to the bus used by the hybrid low voltage system. More details can be
found in~\cite{mastroberardino}.

As with the bias system, single channel control was useful during commissioning
the detector.

\subsection{Hybrid cooling}

The system used to cool the detector readout hybrids is a custom
implementation using closed freon and water circuits. It is designed
to remove energy dissipated by the helix hybrids in the barrel (300 W)
and wheels (120 W). The cooling system is controlled and read out
through a programmable logic device~(PLD), whose serial peripheral
interface (SPI) bus is interfaced via a NIKHEF SPICAN~\cite{SPICAN}
module to CANbus. The PLD continuously monitors environmental
(pressure, temperature, humidity, airflow, etc.)  parameters
associated with the cooling system, and switches off the cooling if an
error threshold is exceeded.  Cold water, $\sim$11.5 $^\circ$C, is
piped to the detector and distributed via a manifold to ladder and
wheel cooling pipes; warm water is returned to the cooling system
input. The entire system is flushed continuously with dry air in order
to avoid condensation. A detailed description of the system including
pictures can be found in~\cite{COOL}.

\subsection{Beampipe and hybrid temperature}

NTC sensors are used to monitor 18 beampipe and 72 hybrid mounted
sensors. The system's SPI bus is interfaced via a NIKHEF SPICAN module
to CANbus. The temperature readout system shares the cooling system's
CANbus.  During data taking the barrel and wheel hybrid temperatures
are typically 18 and 27 $^\circ$C, respectively.

\subsection{Safety interlocks}

The safety interlocks used are designed to ensure that the MVD low and
bias voltages are off during injection or if a hardware failure
occurs. In the off state no power from the MVD slow control system is
dissipated in the detector.  A Frenzel+Berg Easy-30 SPS/CAN
~\cite{FRENZEL}, connected by a separate CANbus, switches off the
power to the hybrid low and bias voltage crates if one of the 4
bi-metal relays ($\sim$60$^\circ$) mounted on the MVD support opens or
if the cooling is off.

Slow control signals are additionally connected to the experiment's
interlock control which prevents beam injection if the MVD is
on. Additional signals disable the trigger if the slow control is
not in the configured state. Note that the experiment's interlock system has no CANbus
interface and the open/close (alarm/no alarm) signals are driven by
logic connected to CST-DI8-TTL and CST-DO8-TTL input/output drivers
~\cite{CSTCAN} connected to a separate CANbus. If the 5 second duty
cycle clock driven from the CST control software stops then injection
and triggers are disabled.

\subsection{Software}

The slow control system reuses the hub based software developed for
the run control system. Each sub-system control task has associated
with it a hub public name, which allows new configuration messages to
be sent and monitoring information to be received.

Configuration messages can be sent from scripts, a ROOT GUI, and via
the Slow Control Control (SCC) system. SCC acts as a scheduler for
slow control operations performed on more than one sub-system.  During
operation the cooling system remains on and the detector is
switched between ON (voltages set) and STANDBY (zero volts) states by
SCC. Additionally, the SCC drives the interlock levels such that
when the detector is not in STANDBY injection is disabled, and when
the detector is neither ON nor in STANDBY the trigger is
disabled.

Monitoring information is made available via html query requests for
tabulated statistics derived from messages held in the hub, or from
histograms showing long term evolution of parameter values. This
information and additional logfile data are archived.

\section{Radiation Protection}
\label{sec:radmon}

Due to the sensitivity towards radiation damage of MVD components, its
close position to the beams and the much increased risks of radiation
damage compared to HERA I running a new radiation monitoring system
was developed~\cite{BLOCH}. The main sources of radiation are the
synchrotron radiation passing through the detector causing possible
damage by direct or back-scattered radiation, off-momentum leptons and
proton beam-gas interactions. In addition partial or total
uncontrolled beam losses are sources of severe damages. The purpose of
the radiation monitoring system is to
\begin{itemize}
\item
monitor the background conditions continously with an
experiment-independent data aquisition;
\item
instantly generate a dump request signal for the lepton machine as
soon as background conditions are too bad.
\end{itemize}
The proton machine operates its own protection system which protects
the experiments as well. The rare uncontrolled losses cannot be
intercepted by dump requests.

\subsection{Radiation sensitive components}

The two most radiation sensitive components of the MVD are the Si
sensors and the CMOS-based readout chips mounted inside the MVD
volume. The sensitivity of the sensors has been tested using hadronic
(reactor neutron irradiations) and ionising radiation ( $^{60}$Co
photon irradiation)~\cite{thesis:dannheim:2003}.

After the neutron irradiation with up to
$\phi_{eq}^\mathrm{max}=1\cdot 10^{13}~n_{eq}$~cm$^{-2}$ (much higher
than the hadronic background expected during the MVD lifetime) no type
inversion showed up as expected. After 3 weeks of annealing the
leakage current had reached a value $420\,{\rm \mu A}$. After the
$^{60}$Co irradiation the leakage current reached 425~$\upmu$A which
reduced to $140\,\upmu$A after 25 days of annealing. Testbeam
measurements showed no change in performance of the sensors in terms
of resolution and reduction of
S/N~\cite{mvdtestbeam:0212037}.

As discussed earlier, the readout is based on the custom designed CMOS
chip HELIX128-3.2~\cite{TRUNK} designed in the AMS CYE
0.8~micron~\cite{AMS} process. One has therefore to expect changes of
threshold voltages and transconductances of the MOS transistors for
irradiations in the 1 kGy range. The design does not use any radiation
tolerant layout techniques. On the other hand the setting of the
transistors of the most sensitive parts of the design can be
controlled by programmable DACs which can compensate the effects of
irradiations to some extent.

\begin{figure*}[hbt!]
\begin{center}
\includegraphics[width=7cm]{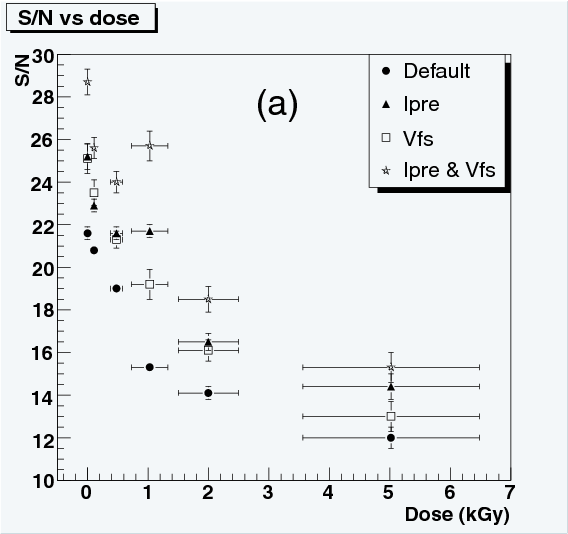}
\hskip0.5cm
\includegraphics[width=7cm]{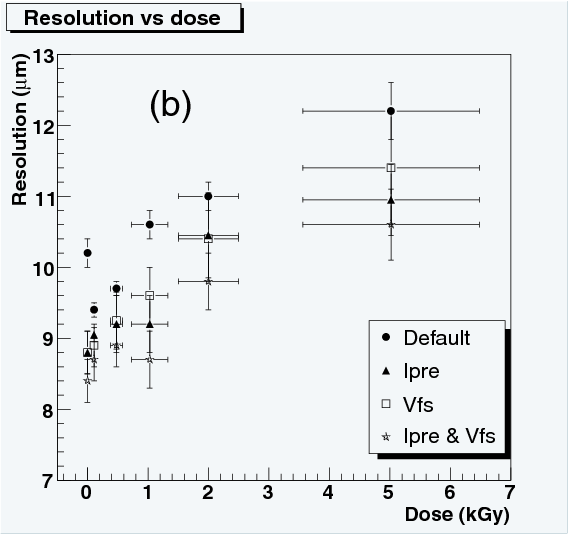}
\caption{S/N-ratio as a function of the accumulated dose as measured
  in the testbeam 
using the default Helix 
settings \emph{(a)}.
Also shown are the S/N-ratios that can be achieved by increasing
$I_{pre}$ to its maximum value, by decreasing $V_{fs}$ to its minimum
  value (just above the threshold voltage)
and by operating both $I_{pre}$ and $V_{fs}$ at their optimum
value. The effect of the 
improvement of the S/N-ratio 
on the resolution is shown in \emph{(b)} where the intrinsic
resolution is 
plotted versus the accumulated dose.}
\label{fig:6.14}
\end{center}
\end{figure*}

For the irradiation of the Helix chips a test module was used
consisting of only one sensor with the hybrid connected which carries
the Helix chips The hybrid was tested in a separate $^{60}$Co photon
irradiation up to 5~kGy~\cite{thesis:velthuis:2003}.
Figure~\ref{fig:6.14} shows the S/N-ratio and the intrinsic position
resolution as a function of the accumulated dose. The setting
determining the bias current of the preamplifier is labelled $I_{pre}$
and the setting determining the feedback resistance of the shaper is
labelled $V_{fs}$.
A deterioration of the performance after irradiation was observed, and
the single track resolution for perpendicular incidence in a single
diode worsens from $7.2\mum$ before irradiation to $12.2\mum$ after
5~kGy. By optimising the programmable readout parameters, the
resolution after irradiation can be improved to $10.6\mum$.  The
resulting performance was thus still within the specified range
\cite{thesis:velthuis:2003}.

In summary both sensitive parts of the MVD can survive a dose of more
than 1~kGy before the performance deteriorates considerably. Since the
MVD was foreseen to operate for five years at ZEUS a dose of more than
20~$\upmu$Gy/s could be safely allowed assuming a year of $10^7$
seconds duty time.

\subsection{Dose rate measurement}
 
\begin{figure*}[hbt!]
\begin{center}
\includegraphics[width=12cm]{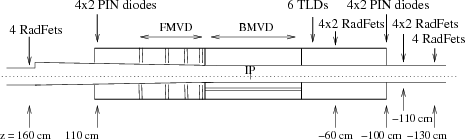}
\vskip0.5cm
\includegraphics[width=7cm]{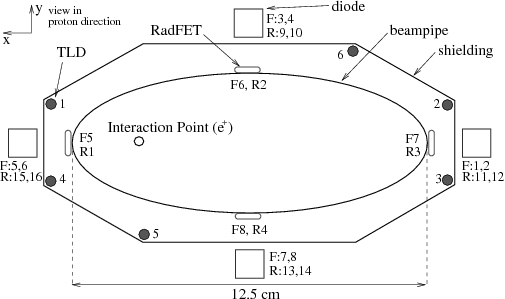}
\caption[~The radiation monitor system inside the ZEUS detector]{The
 radiation monitor system inside the ZEUS detector. a) Position of the
 components along the beam direction. b) Position of the components in
 the $XY$-plane.  The diodes are labeled from 1 to 16 and the
 RadFets from 1 to 8.  The letters F and R in front of the diode and
 RadFet numbers indicate the $Z$ position in the forward ($Z=160$~cm)
 and rear ($Z=-130$~cm) region, respectively. The first diode number
 of each module corresponds to the upper diode (facing the interaction
 point), while the second diode number referes to the diode behind the
 lead shielding. The tubes for inserting the TLDs are labeled from 1
 to 6.  }
\label{fig:radmon-position}
\end{center}
\end{figure*}

The radiation monitor consists of 16 Si pin diodes of $1\,{\rm cm^2}$
size~\cite{sintef} arranged in 8 pairs and mounted inside the MVD
volume as shown in Figure~\ref{fig:radmon-position}.  The two diodes
of each pair are separated by $1\mm$ of Pb to provide a discrimination
between currents produced by mips and photons. Each pair of diodes is
equipped with an NTC resistor to provide a temperature measurement for
leakage current corrections. The diodes are connected via a $30\,$m
doubly shielded cable to the data aquisition and HV
supplies. Initially the diodes have been reverse-biased at
$100\,$V. Later the voltage was reduced to $50\,$V (February 2005) to
allow for more stable running conditions.

The current measured in the reverse-biased diodes consists of the
rate of energy deposited in the diodes and of the leakage
currents. Assuming an energy of $3.6\,$eV for the production of one
electron-hole pair the dose rate and the current are related by

\begin{equation}
\dot{D}~[\mathrm{\frac{Gy}{s}}]=5\cdot10^{-5}\cdot I_\mathrm{photo}
~[\mathrm{nA}].
\end{equation}

The deposition of energy in the diodes leads to radiation damage which
results in an increase of the leakage current. At the time of
writing the original leakage currents of below 1~nA have increased to
about $200\,$nA at $50\,$V biasing for diodes on the inside of the ring.

The currents are fed into a two stage integrating amplifier whose
voltages are measured by a commercial ADC with a resolution of
$10\,$mV/nA. The range of the ADC reaches up to $10\,$V allowing a
maximum current of $1000\,$nA corresponding to $50\,$mGy/s dose
rate. The leakage currents are monitored regularly with no beams in
the storage ring. The measured leakage currents, corrected for
temperature, are the basis of the dose measurements.  For the
monitoring and the control of the dump signals the so called
leaky-bucket concept developed by the BaBar
experiment~\cite{ijmp:a16s1c:1084} is used. This concept allows a
certain amount of excursions of the dose rate. The dump signal is
released only after a certain deposited dose.  Above a threshold of
$0.5\,$Gy/s, the integration of the dose rate starts. A dump signal
will be issued only if a dose of $200\,$mGy has been reached. To
control the dose for rates below the threshold human intervention is
needed.

Initially a software driven system with a $1\,$s sampling rate was
used. In 2006 a new DAQ system was installed which
reduces the reaction times from seconds to milliseconds thereby
reducing the received doses before dumping.

\begin{figure*}[hbt!]
\begin{center}
\includegraphics[width=14cm]{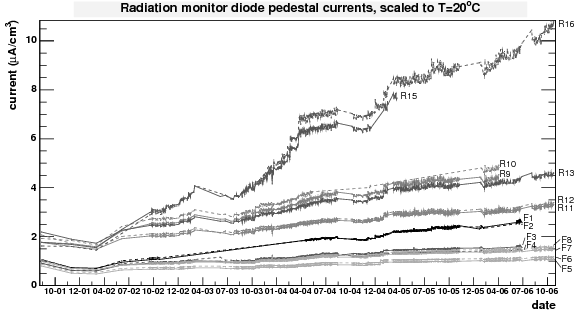}
\vskip0.5cm
\includegraphics[width=14cm]{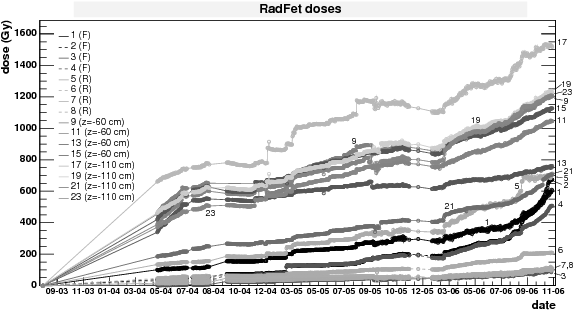}
\caption{Leakage currents of the radmon diodes as a function of date
since October 2001. Please note the scale $\mu$A/$\cm^{3}$. The
normalisation of the leakage current to the depleted volume of the
diodes has been introduced after decreasing the bias voltage from 100
V to 50 V in February 2005.  The lower plot shows the doses measured
by the RadFets as a function of date since September 2003 (2nd
phase).}
\label{fig:radmon-leakage}
\end{center}
\end{figure*}

\subsection{Dose measurement}

In parallel to the current measurement two integrating systems have
been installed at the beginning of the operation of the MVD:
thermoluminescent dosimeters(TLD) and radiation field effect
transistors (RadFet)~\cite{nmrc}. The positions are indicated in
Figure~\ref{fig:radmon-position}.  Whereas the TLDs have to be exchanged
regularly for reading, the RadFets can be read out continously by a
remote data aquisition.  Two types of TLDs have been used:
\begin{itemize}
\item
$^7$LiF (TLD-700) being sensitive to ionising radiation 
\item
and $^6$LiF (TLD-600) being sensitive to both ionising radiation and
thermal neutrons by means of $(n,\alpha)$ capture.
\end{itemize}
In the shutdown in March 2003, 20 months after the beginning of the
HERA II operation, all RadFets were exchanged. In addition another
type of RadFet was installed at $Z=-60$ and $-110\cm$ which can be
read out continously~\cite{nmrc}. The TLDs were not used after March
2003.

\subsection{Measurements and Results}

All dose rate and dose related measurements show a very non-uniform
distribution.  The highest values are found at the ring inside at
negative $Z$, supporting simulations which predict a high rate of
off-momentum leptons (Figure~\ref{fig:radmon-leakage}). The trigger
rates of the ZEUS experiment show the same dependency.  The total dose
received at $Z=-130\,$cm for the ring inside position is approximately
$2\,$kGy. The doses measured with the TLDs until 2003 are in general
agreement with the RadFet measurements. The diodes at $Z=-60\,$cm and
$Z=-110\,$cm which were installed later, received $1.5\,$kGy and
$1.7\,$kGy respectively. During the same period the diodes at
$Z=-130\,$cm received $0.75\,$kGy. Between the two positions $Z=-110$
and $-130\,$cm a scintillator-lead sandwich at $Z=-120\,$cm and vacuum
flanges act as an extra collimator and can therefore explain the
different dose values.

The performance of the MVD modules shows a similar tendency
as the dose measurements: a small increase of noise for modules in
cylinder 0 at the ring inside positions and negative $Z$-position
(Figure~\ref{fig:signoise}) decreasing from negative $Z$ to positive
$Z$ position leading to a decrease of S/N of up to $10\,$\%. A total
dose of close to $3\,$kGy as measured by the RadFets would have led to
a decrease of the S/N of about $33\,$\%. The difference can be
understood by the fact that the Radfets are positioned as close as
$25\,$mm to the beam whereas the readout electronics of cylinder
0/ladder 0 has a distance of close to $50\,$mm from the beam.

\section{Data Quality Monitoring}
\label{sec:dqm}

A three level data quality monitoring system has been developed for
the ZEUS MVD in order to ensure an optimal level of data integrity
under all running conditions. At the first level a fraction of the
data are inspected and relatively simple quantities like occupancy
histograms, hit maps and signal distributions are quickly
calculated. These quantities are constantly monitored by the ZEUS
shift crew. The second level monitoring is tied to the archiving step of
the data: All data that pass the Third Level Trigger are inspected
just after they have been written to the ZEUS offline archive. At this
stage every strip in the detector is inspected and classified. Any
strip whose behaviour in terms of noise or average occupancy deviates
from the expected normal behaviour is classified according to one of
four categories. 
\begin{figure*}[hbt!]
  \begin{center}
  \includegraphics[width=12cm]{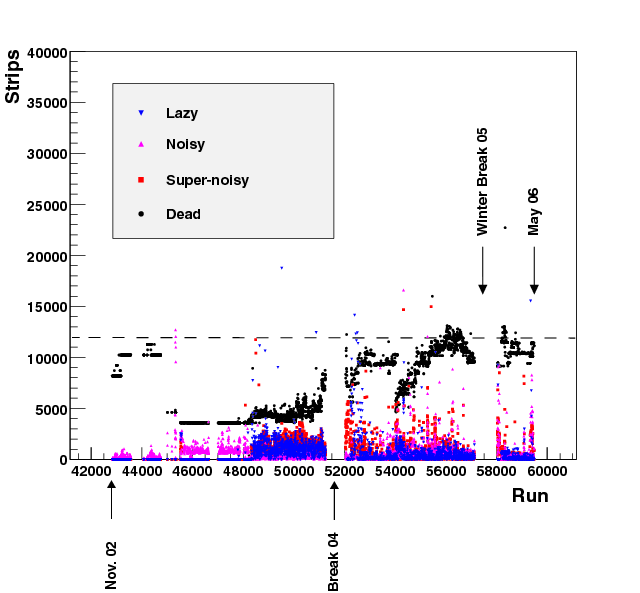}
  \end{center} 
   \caption{Number of flagged channels of the ZEUS MVD as
     a function of the run number. The run range spans the period from
     November 2002 to May 2006. The dashed horizontal line corresponds
     to approximately $5\pc$ dead channels} 
   \label{fig:flagged}
\end{figure*}
These are:
\begin{description}
\item[\em Lazy:] The occupancy of the strip is significantly below the
expectation as calculated from an average over a set of neighbouring
strips.
\item[\em Noisy:] The noise is outside 2.8 standard deviations of the
expectation.
\item[\em Super-noisy:] The noise is outside 3.5 standard deviations of the
expectation
\item[\em Dead:] There are no hits.
\end{description}
Using this classification, channels marked as ``super-noisy'' are
excluded during the track reconstruction in order to simplify the
pattern recognition step.

Fig.~\ref{fig:flagged} shows the number of differently flagged
channels as a function of run number over a period of almost four
years. It can be seen that after an initial rise the number of flagged
channels now fluctuates up and down below a level of ca. 12000 with no
obvious tendency. This level corresponds to approximately $5\pc$ of
all channels of the detector. The origin of these malfunctions is in
the frontend of the detector, the hybrids and cables. These are not
accessible from the outside without a major disassembly of the ZEUS
detector. These malfunctions are not completely understood. However,
the observed failure pattern, namely the fact that typically entire
half modules fail, indicates that the problem is inside the flex PCB
connecting the hybrids to the COMBO cables. This failure mode occurs
almost exclusively in the barrel region (8 failed modules and 6 failed
half-modules in the barrel compared to only 2 failed wheel sectors) It
is suspected that the presence of hygroscopic glue which was used in
the barrel but not in the wheels may have destroyed some of the vias.

\begin{figure}[hbt!]
  \begin{center}
  \includegraphics[width=8cm]{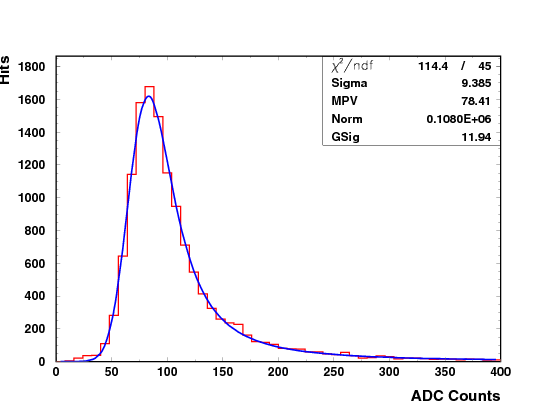}
  \end{center} 
   \caption{Signal in ADC counts for the ZEUS MVD. The fit shown is a
            Landau distribution folded with a Gaussian.} 
   \label{fig:signal}
\end{figure}

The third level of DQM occurs after track reconstruction inside the
reconstruction program. Tracks are used to filter hits which are
associated with charged particles originating from the interaction
vertex. This hit sample is very clean. In addition the association of
hits with tracks allows one to normalize the signals to track length
inside the silicon. Fig.~\ref{fig:signal} shows signals for these hits
corrected for track length. This distribution is well described by a
Landau function with a most probable value around 78 ADC counts
convoluted with a Gaussian. The latter takes care of the simultaneous
presence of signal and noise for the hits.

\begin{figure*}[hbt!]
  \begin{center}
  \includegraphics[width=15cm]{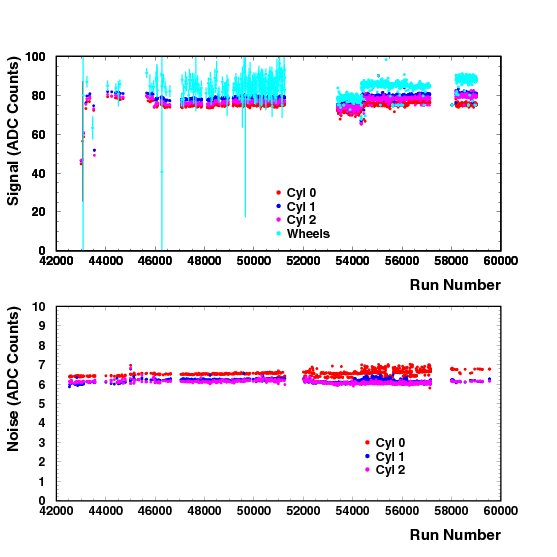}
  \end{center} 
   \caption{Signal and noise as a function of run number in the ZEUS MVD.} 
   \label{fig:signoise}
\end{figure*}
Figs.~\ref{fig:signoise}(a) and (b) show the behaviour of signal and
noise over time, respectively. With the exception of a few jumps in
the signal level, both signal and noise are stable with time. The
jumps occur following a calibration of the sampling delay in the ADC
system relative to the maximum position of the signal pulse.

\section{Alignment and Tracking}
\label{sec:perf}

\subsection{Tracking} 
The offline track reconstruction beyond the local coordinate level
consists of the track pattern recognition, the track fit, the vertex
finding and the vertex fit. The track pattern recognition makes full
use of the track finding powers of the individual detectors and
proceeds through a complex multipass algorithm which combines CTD,
forward and barrel MVD in a transparent way. This has the advantage
that the track finding efficiency of the CTD is effectively enhanced
by the MVD. Seeds are constructed from superlayers of both systems,
and the resulting initial trajectories are then used to collect
additional hits. In the MVD, also seeds consisting partly of barrel
and partly of forward MVD hits are constructed to cover the transition
area. Track candidates with ample hit information from both CTD and
MVD take precedence, but also tracks composed entirely from MVD or
CTD hits are found in later passes of the procedure, which extends the
range beyond the CTD acceptance. The information assigned to a track
then enters the track fit~\cite{mvd-tracking:maddox}, which is based on the Kalman
filter method. The track fit determines the accurate track parameters
and their covariances taking the detailed material distribution of the
MVD and the beampipe elements into account. Effects from both
multiple scattering and ionization energy loss are accounted for. The
track fit also rejects outlier hits based on their $\chi^2$
contribution to the fit. Also the vertex finder and vertex fit make
use of the Kalman filter technique.

\subsection{Alignment}
For the performance in highly resolution-dependent analyses, the
quality of the detector alignment plays a crucial r\^ole. For this
reason a four-step strategy was adopted. In the first step, during
construction, both the position of half modules on the ladders and the
position of the ladders inside the completed detector were surveyed
with a precision of $10\,\mum$. For the second step a laser alignment
system has been incorporated in the overall support structure to
monitor the stability of the MVD in its operating environment. The
primary aim of the system is to define periods of the stability for
the more precise track-based alignment procedures. Those were
performed using both cosmics and electron-proton collisions. In the
following, the methods and results of the MVD alignment procedures are
described in more detail.

\subsubsection{Survey Measurements}

During the assembly three different precision steps were defined:
\begin{itemize}
\item
Gluing of the sensors into a half module; this was performed with a
precision of $5\mum$;
\item
The positioning of the modules on a ladder with a precision of
25~$\mu$m; directly after mounting the positions were measured with a
precision of $5\mum$;
\item
The position of the ladders and wheels in the overall support tube was
determined by the placement of a ceramic ball in a ceramic
cylinder. The requirements were $50\mum$ precision, but after assembly
the position was measured on a 3D survey machine with a precision of
$10\mum$.
\end{itemize}

\begin{figure*}
\begin{center}
\includegraphics[width=0.8\textwidth]{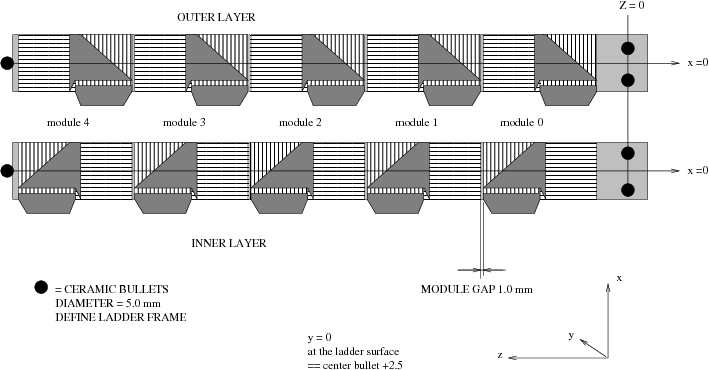}
\end{center}
\caption{Definition of the ladder geometry for survey measurements}
\label{geometry-survey}
\end{figure*}

\begin{figure}
\begin{center}
\includegraphics[width=0.5\textwidth]{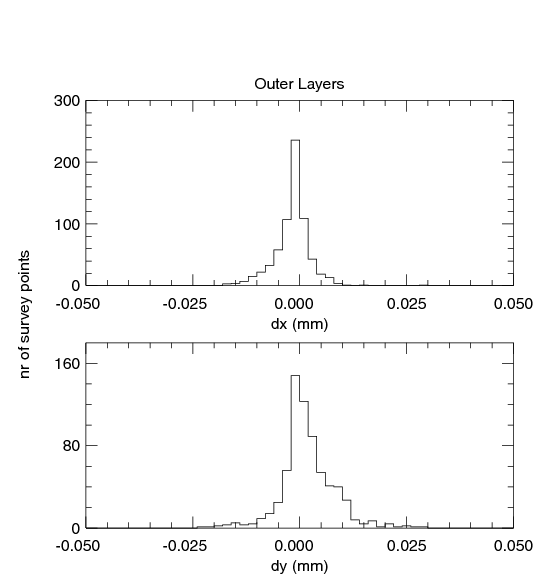}
\end{center}
\caption{Result of the survey positions of the modules in the $XY$-plane as 
compared to the nominal values}
\label{survey-xy}
\end{figure}

Figure~\ref{geometry-survey} defines the coordinates of the ladder for
the survey measurements after the modules were mounted on the ladder
support.  Figure~\ref{survey-xy} summarizes the results of the
position measurements of the modules on the ladders. The differences
between the measured values of the marker positions as compared to the
nominal values are plotted. All points are within $25\mum$ of the
nominal position.

\begin{figure*}
\begin{center}
\includegraphics[width=0.5\textwidth,angle=-90]{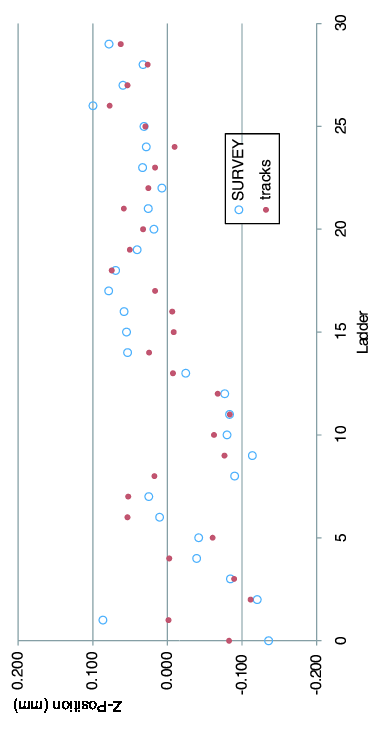}
\end{center}
\caption{Result of the survey of the $Z$-positions of the ladders as 
compared to the nominal values}
\label{survey-z}
\end{figure*}

The $Z$-position of each ladder was surveyed after they were mounted
in the overall support cylinder. The results are compared with cosmic
ray measurements made prior to the installation in the experiment.
Figure~\ref{survey-z} shows that the mounting was precise within
$100\,{\rm \mum}$.

\subsubsection{Laser Alignment}

\begin{figure*}
    \begin{center}
        \includegraphics[width=\textwidth]{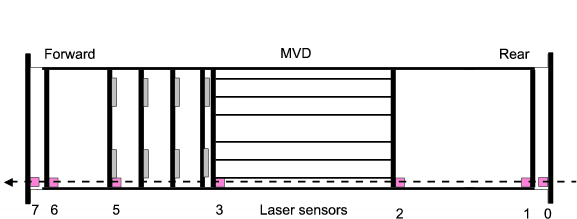}
        \caption{Schematic diagram of one laser alignment beam and sensors. Forward 
and Rear refer to the orientation of the tracking detectors, with forward in the
direction of the HERA proton beam. The sensor numbering is also shown.}
        \label{fig:la-scheme}
    \end{center}
\end{figure*}

The laser alignment system consists of four straightness 
monitors placed around the perimeter of the MVD support tube. Each straightness 
monitor consists of a collimated laser beam, approximately parallel to the 
collider beam line, with seven semi-transparent silicon sensors positioned along 
its path, as shown schematically in Figure~\ref{fig:la-scheme}. Two of the sensors, 
mounted on the forward and rear CTD end-plates (positions 7 and 0), define the line 
for that laser beam. The five remaining sensors are mounted within the MVD at the 
forward and rear end-flanges, the forward and rear barrel flanges and the support 
structure for wheel 3. The alignment system is sensitive to motion of the support 
structure in directions perpendicular to the beam line. Full details of the 
system and its performance are given in \cite{mvd-tracking:mvdla2}.

The laser alignment system measures the residuals at the five sensor
positions along each of the four operational laser lines. A residual
is defined as the distance between the measured laser beam position
and the expected position based on a straight line defined by the two
sensors attached to the CTD.  Two residuals are measured in orthogonal
directions in the plane of the sensor surface in a local coordinate
system.  The most useful information is obtained if a sequence of
laser runs is taken over a long period, with the laser powered
continuously and the beam controlled by a mechanical shutter. Using
this procedure it has been possible to investigate whether, and by how
much, the MVD support structure might move when environmental
conditions change. The most important of these concern the HERA
machine magnets nearest the ZEUS interaction point and the MVD
on-detector electronics. The MVD is supported from the ZEUS CTD and
the support structure has no direct mechanical connection to the
beampipe. However all the service and readout cables - all exiting
from the rear of the detector - are wrapped very tightly around the
final magnet (GG) of HERA. This is a superconducting magnet that
lies inside the beam-hole of the ZEUS RCAL (rear calorimeter) and was
installed as part of the HERA-II upgrade programme. The space between
the magnet and the RCAL modules is very limited, which is why the
cables have to be very tightly bound to the magnet. It is known from
position sensors on the magnet that it moves slightly when the magnet
current is turned on or off. Thus it is possible that such movement
could be transmitted to the MVD, though it is by no means obvious what
type of movement of the MVD might result.

\begin{figure*}
\begin{center}
    \begin{tabular}{cc}
     	\includegraphics[width=0.45\textwidth]{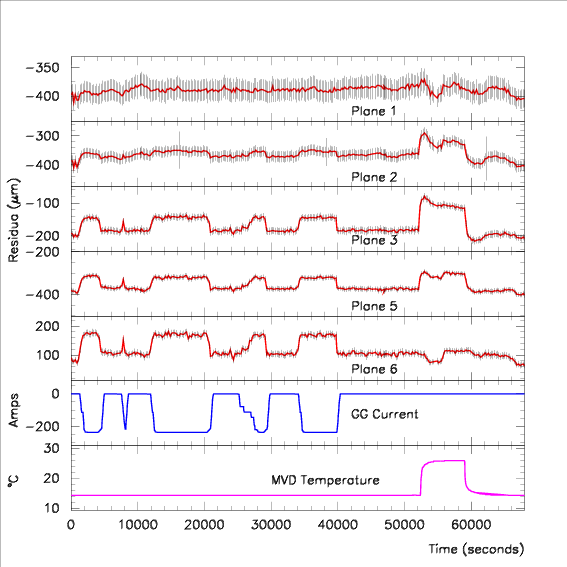} &
        \includegraphics[width=0.45\textwidth]{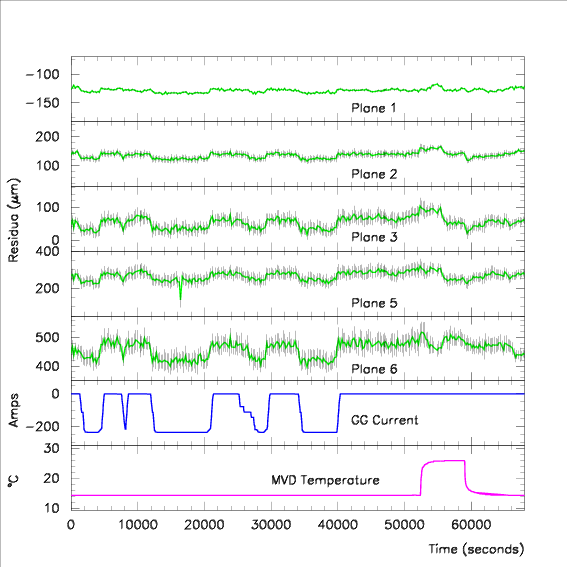}
    	\end{tabular}
        \caption{Beam 2: local $y$ (top), local $x$ (bottom). Residuals for sensor
planes within the MVD as indicated, together with the GG magnet current 
and MVD temperature near wheel3, versus time. }
        \label{fig:beam2}
	\end{center}
\end{figure*}

Figure~\ref{fig:beam2} shows the
residuals for sensor local-$y$ coordinates (left plots) and local-$x$ (right plots) 
for the five sensors within the MVD along laser-beam 2 as functions of
time. The lowest two traces show the magnet current and the MVD temperature 
(measured at a position near wheel 3) for the same period.
Considering the magnet current first, between times of 0 and 45000 there is a 
clear correlation in time 
between the mean residual values themselves and with the magnet current being zero or
non-zero. There is also a tendency for the size of the movement to increase with 
increasing plane number. Such a pattern of movement is consistent with the MVD
support tube being tilted about a fulcrum near the rear CTD support point by the
movement transmitted from the GG magnet by the cables. 
Similarly, for times between 50000 and 65000, there is a correlation between sensor 
position and temperature. 
The change in temperature occurs when the MVD electronics is switched on or off. 
The pattern of movement is different in detail, the biggest effects are now seen in 
the three sensor planes nearest to the on-detector electronics -- at planes 2, 3 and 
5 (MVD barrel flanges and wheel 3).
   
The residual plots show that the position of a sensor is stable while the external
effect is not changing and that the two positions of stability are
themselves stable and reproducible. This has been investigated in more detail by 
averaging the residuals for periods of stability taken from long laser 
runs~\cite{mvd-tracking:mvdla2}. The results show that the difference in position locally can
be as large as $100\,\mu$m, but that any hysteresis effects are at a level of  
$10\,\mu$m or less. 

A long set of laser runs has been taken during a period of physics `luminosity'
running, when all environmental conditions should be stable - with both the GG
magnet and the MVD electronics powered. The results show that the sensor position
is stable to within an RMS of better than $10\,\mu$m  over a period of 7 hours.
It should be emphasised that all the measurements made with the laser system are of
relative positions.

From these studies the following conclusions may be drawn:
\begin{itemize}
\item
The MVD support structure is very stable and there is no indication of any long term
movement or step changes in position;
\item
When external conditions vary, particularly the GG magnet on or off and the MVD
on-detector electronics being powered or not, the MVD support structure shows
movements locally as large as $100\,\mu$m;
\item
Once the conditions return to the previous state the MVD returns to its previous
position, to within $10\,\mu$m.
\end {itemize}

From these results it is clear that track data
for precision alignment should be collected under the operational conditions
of a `luminosity run'.

\subsubsection{Alignment with Cosmics}

For the bulk of tracks from lepton-proton collisions in the MVD only
six 2-dimensional measurements are available. Therefore alignment of
the detector with these tracks using MVD information alone is
difficult. One way to overcome this problem is to use tracks from
cosmic muons that traverse both halves of the detector. These
typically provide twelve measurements in a very clean environment, but
the method is limited by the fact that only small samples of cosmic
muon tracks are available and that certain regions of the detector,
particularly in the horizontal plane, are only poorly illuminated.  In
order to obtain a reasonable number of tracks, data samples recorded
at different times during the first year of operation were
combined. This is possible since the data from the laser alignment
system indicate the position of the detector as a whole to be stable
relative to the global ZEUS coordinate system with a precision of
approximately $10\mum$.  Thus, this method provides a robust initial
determination of the alignment parameters of the detector. A detailed
description of the method used and of the results can be found
in~\cite{kohno}

An iterative two step method was used both to determine the global
displacement of the entire MVD relative to the global ZEUS coordinate
system as well as to align MVD ladders relative to each
other. Statistics were not sufficient to perform an alignment of
individual half modules with the cosmics data. In the first step well
measured tracks were refitted after removing individual hits. In the
second step the residuals of the ``removed'' hits were used in a
least-squares fit in order to determine translational and angular
alignment parameters.  For the global alignment 6 free parameters,
namely 3 translations and three angles, were used in the fit. For the
local alignment a total of 180 parameters were used, namely the
positions and angles of the 30 ladders in the barrel part of the
detector. The global alignment was iterated 10 times, followed by 5
iterations of the local alignment. The whole procedure was then
repeated. The method converged quickly and terminated after
two global and three local iteration steps. The alignment with cosmics
was not used to align the forward part of the MVD due to lack of
coverage.

It should be noted that with this method track and alignment
parameters are fitted independently of each other thus neglecting any
correlations between them. Nevertheless, the method yields good robust
results. 
\begin{figure*}[htbp!]
\begin{center}
\includegraphics[width=15cm]{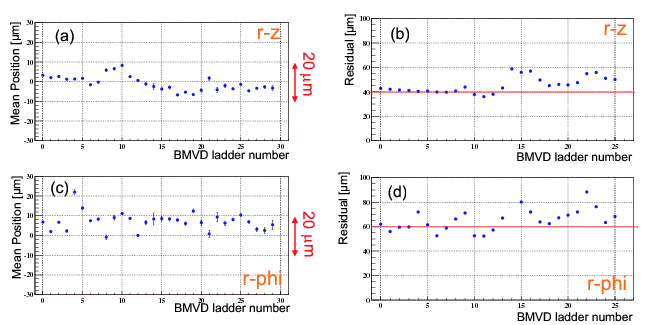}
\end{center}
\caption{Track residuals, mean position~(a) and standard
      deviation~(b) in $R\phi$ and mean position~(c) and standard
      deviation~(d) in $RZ$, after the cosmic alignment procedure.}
\label{track:cosmresid}
\end{figure*}
This can be seen in Fig.~\ref{track:cosmresid} which shows (a) the
mean and (b) the width of track residuals for tracks from cosmic muons
for the thirty ladders in the BMVD after applying the alignment
procedure. These numbers can be considered as figures of merit for the
cosmic alignment: The means are smaller than $10\mum$ and the
widths are about $60\mum$ and $80\mum$ for the
$RZ$ and $R\phi$ sensors respectively.

\subsubsection{Alignment with tracks from ep Collisions}
The final alignment was done using tracks from electron-proton
collisions.  While cosmic muon events have very clean signatures that
lend themselves ideally for alignment purposes, they do not illuminate
the full phase space, and are ultimately limited by statistics. The
best alignment with cosmics can be expected for the barrel modules in
the upper and lower parts of the detector, while the sideways barrel
sensors are already less accessible to alignment with cosmic
muons. The forward MVD is entirely inaccessible with this kind of
alignment.

The largest sample of tracks for alignment is available from regular
ep collisions. With 150 million events recorded in a good HERA-II
year, such tracks are available in abundance even under very
restricted selections, and they illuminate all detectors to the same
degree as they are used for physics. Another advantage is that these
alignment tracks are taken under identical conditions as for physics
runs.

In comparison to cosmic muons, tracks from ep collision tracks start
in the middle of the detector, hence in the barrel they traverse
typically only six measurement layers of silicon, which
leads to three hits in $R-\phi$ and three hits in $Z$ sensors, though
higher or smaller numbers of hits are possible depending on the
trajectory. The transverse track parameters curvature, azimuth angle
and impact parameter can obviously not be overconstrained by three MVD
$R-\phi$ hits. To improve the constraints on the trajectory, two additional
pieces of information are included:
\begin{itemize}
\item the parameters of the CTD segment of the track helps
to pin down the momentum of the track.
\item the run-by-run beam spot gives an additional constraint near the
middle of the MVD.
\end{itemize}

Due to the relatively small number of MVD hits on $ep$ collision
tracks, it would not make sense to compute an unbiased residual by
refitting the track with the hit of the layer in question excluded, as
it is done in the alignment procedure for cosmic muons. Therefore, the
residuals of a hit will be biased by the fact that hit coordinate and
track parameters are correlated. This will not diminish the amount of
information reflected in the hit residual, but the alignment procedure
must take this correlation into account. It is therefore necessary to
perform a global fit of all alignment parameters and all tracks in the
alignment sample simultaneously. Since the aim is the individual
determination of the positions and rotation angles of 600 barrel
sensors with about 1 million tracks, this fit contains a large number
of free parameters: about 3000 alignment parameters and several
million track parameters. A specialized engine for this kind of fit is
available called Millepede algorithm~\cite{mvd-tracking:millepede}. We
have used it within the $ep$ collision-based alignment.

\begin{figure*}[htbp!]
\begin{center}
\includegraphics[width=12cm]{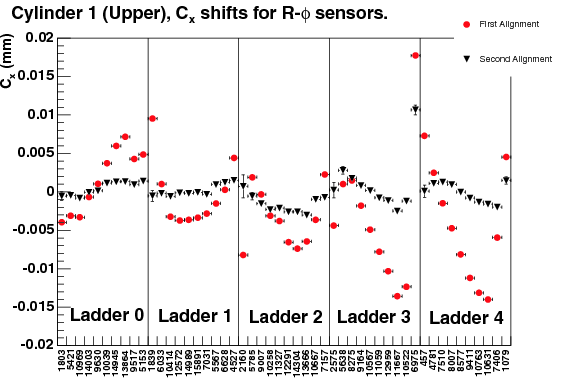}
\end{center}
\caption{Alignment constants describing the offset in measurement
direction for the $R\phi$ sensors of five ladders in the upper middle
cylinder. The circles represent the constants determined by the ep
collision track alignment procedure applied on the cosmics-aligned
geometry. The numbers at the bottom are the numbers of tracks used in
the alignment of each sensor. The triangles are the constants
determined in a second pass of ep collision track alignment relative
to the first pass.}
\label{epconstantscx}
\end{figure*}
A small subset of the resulting constants is shown in
Figure~\ref{epconstantscx}. The resulting shifts in the
measurement direction of the sensor, $C_x$, for the $R-\phi$ sensors
of the five ladders in the middle cylinder in the upper half of the
MVD are displayed. Within each of these ladders, the sensors are displayed in the
order of their position along the proton beam direction. Most error
bars are smaller than the symbol size, but one should be aware
that the millepede fit does not account for multiple scattering, and
hence the real errors will be larger. The alignment parameters range
typically between $\pm 150\mum$. The strong correlation of
neighboring sensors certifies to the high accuracy of the MVD assembly
and the microscopic survey within each ladder. The dominantly linear
variations of alignment parameters within the ladders indicate
rotation around an axis normal to the ladder plane. The rotation
angles are small ($\ll 1\,$mrad) and become visible only due to the
large leverage; they may reflect both adjustments of the ladder
positions within the MVD frame, as well as a global rotation of the
MVD relative to the nominal beam line. For some ladders the alignment
constants show a non-linear pattern, one may speculate that their
origin stems from a subtle deformation of the ladder, or from a
warp-like buckling of the sensors.

\subsubsection{Testing alignment performance}
After implementing the alignment parameters in the MVD geometry, the
consistency of the alignment procedure has been tested by reprocessing
the track reconstruction with the updated geometry, and repeating the
alignment procedure on the reprocessed sample. This second pass would
also indicate a possible need for iteration, which might arise in case
pattern recognition has been seriously suffering from initial
misalignment. The result of the second alignment pass is shown by the
triangular symbols in Figure~\ref{epconstantscx}. Since the resulting
constants are close to zero, one concludes that the alignment
procedure is self-consistent, and iteration is not necessary. The
remaining variance of the points indicates a typical internal
alignment accuracy of approximately $25\mum$ in measurement direction.

\begin{figure*}[htbp!]
\begin{center}
\begin{tabular}{cc}
\includegraphics[width=8cm]{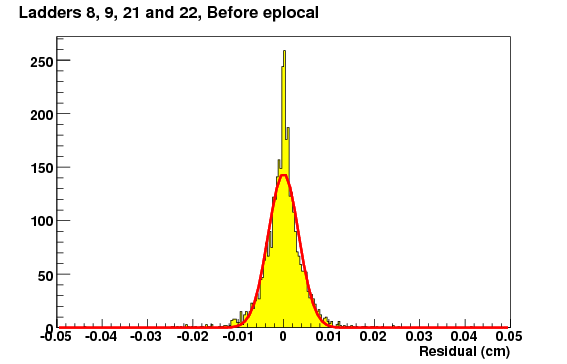} &
\includegraphics[width=8cm]{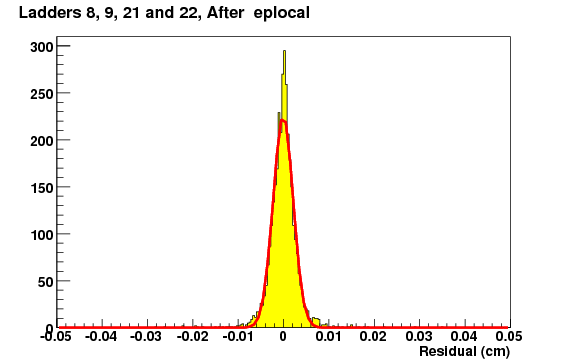}
\end{tabular}
\end{center}
\caption{Distribution of residuals of four ladders near $\varphi 
\approx 180^{\circ}$ before (left) and after (right) alignment with tracks 
from ep collisions. The fitted curve is a Gaussian with a width of
$34\mum$ (left) and $22\mum$ (right).}
\label{epresiduals}
\end{figure*}
The improvement of the track quality by the alignment step is
directly reflected in the distribution of the residuals. In
Figure~\ref{epresiduals}, the residual distributions for hits in the
four ladders near $\varphi
\approx 180^{\circ}$ are shown before and after alignment with $ep$
collision tracks. The fitted width is found to improve considerably from
$34\mum$ to $22\mum$.

\begin{figure*}[htbp!]
\begin{center}
\includegraphics[width=15cm]{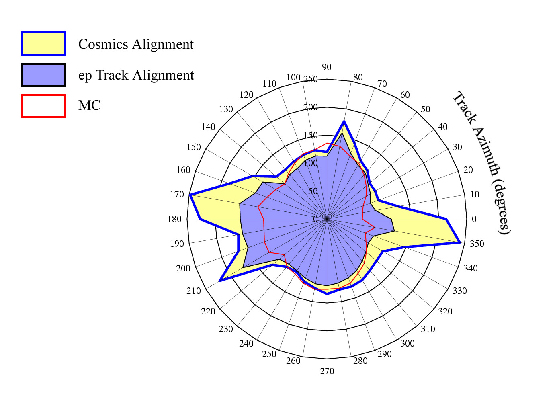}
\end{center}
\caption{Polar diagram showing the visible impact parameter resolution
(in $\mum$) with respect to the beam spot as radius as a function of
the track azimuth angle $\varphi$, for tracks with $p_T>3~GeV$ having
at least 2 hits both in the $r-\varphi$ and the $Z$ sensors. This
resolution includes the intrinsic size of the interaction region, and
is shown as the radial coordinate in units of $\mum$. The
light-shaded outer contour displays the impact parameter resolution at
the level of alignment with cosmic muons, the darker-shaded inner
contour is obtained after the additional alignment step with tracks
from ep collisions.  The thin line without shade is the
corresponding MC result, which does not simulate misalignment.}
\label{radar}
\end{figure*}
It is very important to find means of monitoring the track quality
beyond simple residuals, and ideally they should be independent of the
vertex reconstruction algorithm since the latter introduces an
additional selection bias. A very useful observable is the impact
parameter of high transverse momentum tracks with respect to the beam
spot. The vast majority of high $p_T$ tracks in $ep$ collisions at
HERA are produced at the interaction point, within experimental
resolution. The visible impact parameter distribution is therefore
characterized by a narrow peak centered around zero, while decay particles
from long-lived states only appear as a shallow background. The fitted
width of the peak thus reflects the track resolution including
effects from residual misalignment, plus a contribution from the beam
spot width.

The fitted width of the visible impact parameter as a function of
azimuth angle $\varphi$ is shown in fig.~\ref{radar}, for tracks with
$p_T>3~GeV$. The coordinates of the beam spot have been determined by
averaging primary vertex positions at run level. At the level of
alignment with cosmic muons, the resolution is already close to the MC
expectation in much of the upper and lower areas of the barrel, while
it is relatively poor -- up to $250\mum$ -- in the azimuth range
between 160$^\circ$ and 210$^\circ$, and between 350$^\circ$ and
0$^\circ$. This effect is not very surprising, since these regions are
expected to be least covered by the cosmic muon sample. The alignment
with tracks from $ep$ collisions is found to improve the impact
parameter resolution in all azimuth regions, and the improvement is
particularly large in the problematic areas.

\subsection{Tracking efficiency}

\begin{figure*}[htbp!]
\begin{center}
\includegraphics[width=15cm]{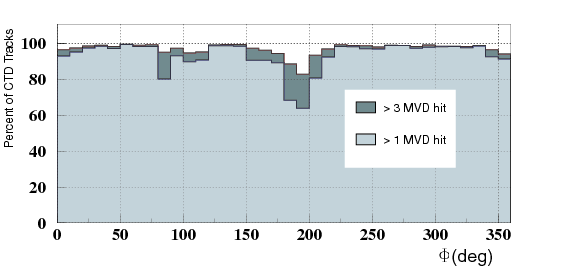}
\end{center}
\caption{An estimate of the tracking efficiency of the ZEUS MVD. The percentage of tracks found by the CTD in the fiducial volume
of the MVD that have 2 and 4 associated hits in the MVD respectively
are shown by the light and dark grey histograms.}
\label{trackeffic}
\end{figure*}

The tracking efficiency is a quantity that depends not only on details
of the detector but to a significant degree on details of hit and
track reconstruction, such as clustering algorithms, track pattern
recognition, the track fit, the alignment of the detector and many
other parameters. Here we give only a simple estimation of this
quantity which indicates that the tracking efficiency is indeed
high. For this purpose Figure~\ref{trackeffic} shows the percentage of
tracks found in the CTD that have either at least 2 or at least 4
associated hits in the MVD as a function of the azimutal angle.
This quantity approximately measures the MVD efficiency folded with
that of the track pattern recognition. Impurity in the CTD hit
selection causes a slight underestimate while impurity in the MVD
cluster selection causes a slight overestimate.  Overall the
efficiency is flat at a level in excess of $98\,\%$ as a function of
the polar angle with three characteristic dips near $0^\circ$,
$80^\circ$ and $180^\circ$. These dips are due on the one hand to
holes in the geometrical acceptance of the detector and to an
accumulation of bad channels. In particular these are two small holes
in the coverage of the inner cylinder near $0^\circ$ and $80^\circ$,
the presence of only two cylinders near $180^\circ$ and an
accumulation of bad channels around $80^\circ$.

\subsection{Specific energy loss}

\begin{figure}[htbp!]
\begin{center}
\includegraphics[width=0.45\textwidth]{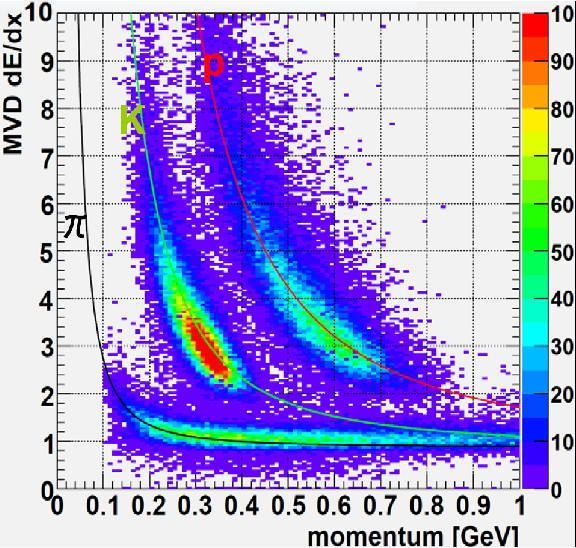}
\end{center}
\caption{The specific energy loss for normal incidence of tracks in the sensitive volume of
the MVD as a function of the track momentum. The specific energy loss
has been calibrated to one for a pure sample of pions from neutral
kaon decays.}
\label{fig:dedx}
\end{figure}

The analog detector electronics provides a 10-bit value of the
observed pulse height which is related to the specific energy loss of
particles traversing the sensitive volume of the detector. These
values calibrated to one for a pure sample of pions, e.~g. from the
decay of neutral kaons, are shown in Figure~\ref{fig:dedx}. One can
easily identify the three different bands stemming from protons, kaons
and pions. The resolution is around $11\,$\% varying slightly with
momentum and particle species.

\subsection{Impact on physics signals}
\begin{figure*}[htbp!]
\begin{center}
\includegraphics[width=15cm]{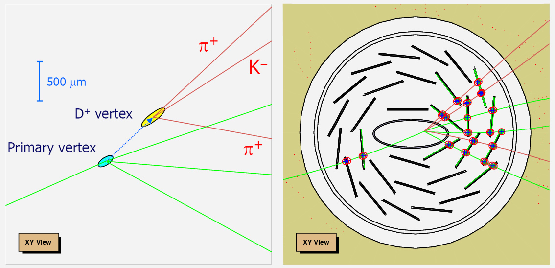}
\end{center}
\caption{Display of an event with a $D^+$ candidate.}
\label{dplevent}
\end{figure*}
\begin{figure}[htbp!]
\begin{center}
\includegraphics[width=7cm]{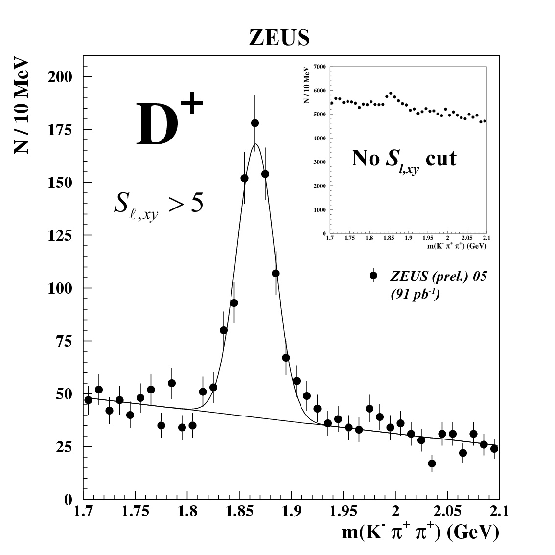}
\end{center}
\caption{Reconstructed mass of $D^+$ candidates.}
\label{dplmass}
\end{figure}
The improved track reconstruction and accuracy of the MVD has enabled
the experiment to use impact parameter and decay length signatures in
heavy flavor analysis. One example is the reconstruction of the $D^+$
meson in its decay channel $D^+
\rightarrow K^-\pi^+\pi^+$. The MVD allows detection of this decay
pattern routinely as a displaced secondary vertex, as illustrated in
the event display in Figure~\ref{dplevent}.  For each accepted $D^+$
candidate, the decay length in the plane transverse to the beam,
$l_{xy}$, has been calculated as
\begin{equation}
l_{xy} = \frac{1}{\sqrt{p_x^2+p_y^2}}
  \left( \begin{tabular}{c} $p_x$ \\ $p_y$ \end{tabular} \right)
  \left( \begin{tabular}{c} $X_{vtx} - X_{bsp}$
                         \\ $Y_{vtx} - Y_{bsp}$ \end{tabular} \right)
\end{equation}
where $p_x$ and $p_y$ are the components of the transverse momentum of
the $D^+$, $X_{vtx}$ and $Y_{vtx}$ are its vertex coordinates, and
$X_{bsp}$ and $Y_{bsp}$ are the transverse coordinates of the beam
spot.  The decay length significance has been computed as $S_{l,xy} =
l_{xy}/\sigma(l_{xy})$, where the error $\sigma(l_{xy})$ takes the
resolutions of tracks and beam spot into account. Fig.~\ref{dplmass}
shows the resulting invariant mass spectrum for $S_{l,xy}>5$ and
$p_T>3$~GeV, for DIS events collected in the $e^-p$ run between
December 2004 and September 2005, corresponding to an integrated
luminosity of $91~pb^{-1}$. The mass spectrum shows a very clean
signal of $D^+$ mesons with very little background, thanks to the
suppression of tracks from the primary interaction. The data sample
has been used to determine the mean lifetime of the $D^+$ with an
accurary of 10\%~\cite{mvd-tracking:dlife} in agreement with the world
average. This illustrates the flavor-tagging capabilities of the ZEUS
Micro-Vertex Detector.

\section{Summary and Conclusions}

In this article we have described the design, construction and
performance of the Microvertex Detector built for the ZEUS experiment
at the HERA collider at DESY. In this final section we briefly review
key design choices and other decisions taken during the construction
of the detector and comment on how the detector has performed as a
consequence of these.

A silicon strip detector was installed in the ZEUS experiment at the
HERA collider at DESY in 2001 and by the end of the HERA running in
2007 an integrated luminosity of $350\,$pb$^{-1}$ with 99\% reliable MVD
data was recorded.  An impact parameter resolution of $100\,\mu$m has
been achieved reaching design specifications.

A 60 cm long barrel with three concentric silicon layers surrounds the
interaction point with layers at radii of approximately 4, 9 and
12~cm. In the forward proton direction four wheels are mounted
perpendicular to the beamline. The total silicon surface amounts to
$2.5\,$m$^2$.

The readout pitch is $120\,\mu$m with 5 intermediate strips enhancing
the charge division.  In a readout module two sensors are joined using
a Upilex cable with copper strips and then connected to the analog
HELIX3.2 chip.  In a testbeam a spatial resolution of $7.2\,\mu$m s been
measured for perpendicular tracks on a single sensor. For a half 
module a resolution of $13\,\mu$m was measured.

Lightweight carbon fiber structures support the sensors, together with
the readout chips and required water cooling and cabling; the average
amount of material is 3\% of a radiation length per layer for
perpendicular incidence.

The signal-to-noise has been stable over the five years of data
taking. The number of dead channels slowly increased to 5\%. The
origin is not quite certain but there is some suspicion about the
stability of the connection of the hybrids with the readout chips to
the sensor and readout cable.

A radiation monitor system was installed since the silicon sensors as
well as the Helix chip are radiation sensitive components.  The
radiation hardness requirement was a total dose of $2\,$kGy, four
times the expected dose.  Three types of monitors were used: pin
diodes, thermoluminescent dosimeters and radiation field effect
transistors.  The worst case estimate is a dose of a few kGy after 6
years of operation resulting in a degradation of the signal\/noise
limited to 10\% for a few sensors at the inner cylinder of the barrel.

During construction the positions of the sensors on a ladder were
carefully checked and archived with optical survey measurements. Small
deviations from design ($\approx 10\,\mu$m) were found and therefore
none of these measurements were used during reconstruction.  Tracks
from cosmic muons provide the alignment of the ladders inside the
overall frame.  A laser alignment system monitors the stability of the
vertex detector.  When the nearby superconducting low beta magnets
were switched on, deformations of the order of $100\,\mu$m were
observed. Under stable running conditions it is however shown that the
geometry is stable within $10\,\mu$m.  At present the alignment accuracy
is below $25\,\mu$m using tracks from collisions.

Kalman filter techniques are used for track fitting as well as vertex
finding.  Apart from a region in $\phi$ where only two layers of
silicon are traversed the impact parameter resolution has reached the
design value of $100\,\mu$m for high momentum tracks.

\section{Acknowledgments}

The authors would like to acknowledge the work and dedication of many
people who made this project possible. Not all can be mentioned
here. However, the many contributions of technicians and engineers
from the collaborating institutes, namely F.~Benotto,
H.~Boer~Rookhuizen, H.~Botschig, P.~de~Groen, T.~Handford,
J.~Hauschildt, J.~Hill, M.~Jaspers, R.~Kluit, H.~Kok, M.~Kraan,
K.~Kretschmer, J.~Kuyt, B.~Payne, U.~Pein, A.~Rietmeiyer A.~Speck,
O.~Strangfeld, P.~P.~Trapani, P.~Verlaat and A.~Zampieri are highly
appreciated.

\providecommand{\etal}{et al.}
\providecommand{\coll}{Coll.}
\providecommand{\nima}{Nucl. Instrum. Methods~A}

\end{document}